\documentclass[a4paper,11pt]{article}
\usepackage{jheppub}
\usepackage{physics}
\usepackage[T1]{fontenc}
\usepackage{lineno}
\usepackage{xcolor}
\usepackage{tikz}
\usepackage{soul}
\usetikzlibrary{decorations.markings}
\usetikzlibrary{arrows.meta,calc,decorations.pathmorphing,patterns}
\usetikzlibrary{decorations.pathmorphing}

\title{\boldmath Kerr/CFT Traversable Wormhole with Fermionic Double-Trace Deformation}

\author[a]{M. Zhahir Djogama,}
\author[a]{Fitria Khairunnisa,}
\author[b]{Hadyan Luthfan Prihadi}
\author[a,c]{and Freddy Permana Zen}

\affiliation[a]{Theoretical High Energy Physics Group, Department of Physics, FMIPA, Institut Teknologi Bandung, Jl. Ganesha 10 Bandung, Indonesia}
\affiliation[b]{Research Center for Quantum Physics, National Research and Innovation Agency (BRIN), South Tangerang 15314, Indonesia}
\affiliation[c]{Indonesia Center for Theoretical and Mathematical Physics (ICTMP), Institut Teknologi Bandung, Jl. Ganesha 10 Bandung,
	40132, Indonesia}

\emailAdd{zhahirdjo@gmail.com}
\emailAdd{30223301@mahasiswa.itb.ac.id}
\emailAdd{hady001@brin.go.id}
\emailAdd{fpzen@fi.itb.ac.id}

\abstract{
    The construction of a traversable wormhole with double-trace deformation has been achieved so far by using boson fields as the perturbation. In this work, we study double-trace deformation with fermion fields in the two-sided Kerr background to open a traversable wormhole. We construct the fermionic double-trace deformation within the Kerr/CFT framework. We consider the near-horizon, near-extremal Kerr geometry, which is dual to a conformal field theory. The lack of fermionic superradiance let us describe the wormhole at every region, even at the off-axis region where bosonic field experiences instability due to superradiance. By choosing a certain coupling between the left and right boundaries, the two-point function is modified, and its first order correction contributes the negative energy to open the wormhole. The wormhole is most traversable when the perturbation is turned on at early times, with opening that depends on the mode's frequency, the black hole temperature, and the fermion mass. At late times, the average null energy has damped oscillation behavior until eventually reaches zero. Wormhole with lower temperature have less traversability and it is completely closed at extreme limit. The same traversability can also be reproduced by using eikonal method. On the other hand, rotation and the number of fields involved in the deformation can increases the upper bound on information transfer up to the order of the entropy. Additionally, symmetrical effective potential bumps connected by the wormhole can produce observable echoes. We find that the echo time delay cannot exceed the scrambling time of the black hole.
}

\begin{document}
\maketitle
\flushbottom

\section{Introduction}\label{sec:intro}
Wormhole is a structure that connects two causally disconnected regions in spacetime. Einstein and Rosen used this concept to formulate a bridge that connects two identical Schwarzschild spacetimes, later known as the Einstein-Rosen bridge \cite{EinsteinRosen}. However, this wormhole quickly closes before a light ray can pass through it, making it non-traversable \cite{VisserPhysRevD.39.3182,FullerW}. The traversability of wormholes is prevented by the average null energy condition (ANEC)
\begin{equation}\label{ANEC}
    \int T_{\mu\nu}k^\mu k^\nu d\lambda \geq 0,
\end{equation}
which implies that the average stress energy tensor along a null geodesic is always non-negative. In order for wormholes to be traversable, ANEC needs to be violated by introducing negative energy. One of the sources of this negative energy is called "exotic matter"\cite{MorrisPhysRevLett.61.1446}. However, it introduces another problem because exotic matter has not been observed in the real world \cite{Lobo2005,Hayward2002,Sushkov2005,Zaslavskii2005}.

Gao, Jafferis, and Wall (GJW) proposed an alternative mechanism for constructing traversable wormholes by introducing a deformation that couples the two asymptotic boundaries non-locally \cite{GaoJW}. They showed that this deformation can generate negative null energy contributions arising from quantum effects. This construction was first realized in the context of the AdS/CFT correspondence \cite{Maldacena1999}, where a double-trace deformation of a two-dimensional boundary CFT is holographically dual to scalar fields propagating in the three dimensional BTZ black hole geometry. The coupling between operators on each boundary deforms the boundary, described as a double-trace deformation, and generates the necessary negative null energy for the wormholes to be traversable. The double-trace deformation modifies the stress energy tensor in form of bulk-to-boundary propagators. These propagators is obtained by using point-splitting method, where one point on the bulk-to-bulk propagator is taken to the boundary. To further support this mechanism, the traversability result from point-splitting method can also be derived by using eikonal method \cite{Maldacena2017, caceres2018, Ahn:2020csv}.

The holographic nature of the traversable wormholes via double-trace deformation suggests that there is a connection between traversable wormholes and quantum teleportation. Indeed, this duality is in line with the famously known ER=EPR relation, where two entangled black holes due to quantum entanglement or Einstein-Podolsky-Rosen correlation (EPR correlation) are connected by a wormhole or ER bridge \cite{Maldacena2013}. In fact, the GJW traversable wormhole is related to a teleportation protocol that involved two entangled systems in thermofield double state \cite{Gao2021Traversable, Jafferis2022Traversable}. In this protocol, after a weak coupling between the two systems, quantum information is teleported from one system to another. This connection and the development of a quantum computer open up the possibility of studying the quantum properties of gravity in a lab. 

The GJW traversable wormhole construction has been extended to a variety of gravitational backgrounds, including two-dimensional black holes \cite{Maldacena2017}, rotating BTZ geometries \cite{caceres2018}, and AdS–Schwarzschild spacetimes \cite{Bak:2018txn}. Another attempt has been made to construct a traversable wormhole from a more physical black holes, the Kerr black holes \cite{Bilotta}. Although Kerr is asymptotically flat and does not have dual to CFT, it turns out that the near-horizon region of extreme Kerr has $SL(2,\mathbb{R})\times U(1)$ symmetry \cite{Guica2009}. It is well known as the Kerr/CFT correspondence. Moreover, for non-extreme Kerr, the conformal symmetry is hidden in the near-horizon region \cite{Castro2010}. With proper limits, it has two timelike boundaries, unlike extreme Kerr, that can be exploited to construct a traversable wormhole. However, bosonic fields in Kerr geometries are subject to superradiant amplification \cite{DeWitt:1975ys}, which complicates the global definition of traversability, particularly in off-axis regions \cite{ottewill2000}. 

Most existing realizations rely on bosonic operators to induce double-trace deformation, with generalizations to vector \cite{Ahn} and tensor fields \cite{khairunnisa2025} also explored. However, a systematic understanding of holographic traversability requires us to examine how the spin and statistics of the mediating field influence the mechanism. In particular, fermionic fields introduce qualitatively new features compared to bosons. The choice of boundary conditions for spinor fields is more subtle \cite{HENNINGSON199863, Muck1998, Henneaux:1998ch}, since the on-shell bulk action vanishes and the dynamics are entirely governed by boundary terms dictated by the variational principle. Different boundary terms can lead to different conformal field theories, such as Lorentz violating boundary terms \cite{Laia2011AHF}. Extending traversable wormholes with fermionic double-trace deformation could provide another perspective for the holographic teleportation and quantum gravitational phenomena.

Moreover, the distinction between bosonic and fermionic statistics has direct physical consequences in rotating backgrounds. The absence of fermionic superradiance allows for a Hartle–Hawking-type vacuum state that remains regular in the off-axis region \cite{Casals2013}, provides a consistent and globally well defined construction throughout the near-horizon near-extremal Kerr geometry. In this sense, fermionic double-trace deformations provide a more robust mechanism for generating traversable wormholes, especially in stationary spacetime. In addition, rotating wormholes can potentially have a higher upper bound on information transfer \cite{caceres2018}. However, in the rotating BTZ case, rotation alone is not enough to increase the amount of information transferred from the order $r_H/l$ to the order of entropy $S_\text{BH}\sim r_H/G_\text{N}$, which is the theoretical maximum \cite{Freivogel2019}.  Our analysis clarifies how spin, angular momentum, and fermion mass, control holographic teleportation and wormhole traversability in rotating black holes.

Even with a robust mechanism for the construction of wormholes, observing the wormholes remains a challenge. Fortunately, a well-defined description of a rotating physical wormhole opens up the possibility of a unique observable. Two different regions of Kerr connected by the wormholes create a symmetric potential system \cite{Bueno2018}. The double potential bump arises from the fact that each side of the Kerr geometry has a characteristic potential barrier. Incoming signals can be trapped inside this symmetric potential and leak out as echoes that can be observed. The time delay between echoes is proportional to distance travel by the signal from one potential to the other, or in this case proportional to the wormhole throat. Thus, we can calculate the echoes time delay from our calculation of the wormhole opening with fermionic double-trace deformation.

This paper is organized as follows. In Section \ref{sec:gravitysetup}, we review near-horizon, near-extreme Kerr geometry as our gravity setup, and the Dirac equation in this geometry. In Section \ref{sec:fermionicdt}, we construct fermionic double-trace deformation in Kerr by first evaluating the Dirac action at the boundary. In Section \ref{sec:wh}, we apply the deformation to modify the two-point function that appears in the expectation value of stress energy tensor. We calculate the average null energy to see the traversability of the wormholes, the bound on information transfer, and the time delay between echoes that the wormholes produce. To verify the result, we also derive the traversability by using eikonal method. In Section \ref{sec:discuss}, we summarize the results and present the discussion.

\section{Near-NHEK Geometry}\label{sec:gravitysetup}
\subsection{Near-Horizon Near-extreme Kerr Metric}
In this section, we summarize the background geometry where the wormhole will be constructed. The near horizon geometry of a near extreme Kerr black hole (nNHEK) is constructed by taking certain limits of the Kerr black hole. For a rotating black hole with angular momentum $J$ and mass $M$ ($J=aM$), the Kerr metric in the Boyer-Lindquist coordinate can be written as
\begin{align}\label{kerrmetric}
	ds^2 = &-\left(1-\frac{2M\hat{r}}{\hat{\rho}^2}\right)d\hat{t}^2
	+ \left(\hat{r}^2+a^2+\frac{2a^2M\hat{r}\sin^2\theta}{\hat{\rho}^2}\right)\sin^2\theta d\hat{\phi}^2
	-\frac{4aM\hat{r}\sin^2\theta}{\hat{\rho}^2}d\hat{\phi} d\hat{t} \nonumber\\
    &+ \frac{\hat{\rho}^2}{\Delta}d\hat{r}^2 +\hat{\rho}^2d\theta^2,
\end{align}
where
\begin{equation}
	\hat{\rho}^2 = \hat{r}^2 + a^2\cos^2\theta,~~~
	\Delta = \hat{r}^2+a^2-2M\hat{r}.
\end{equation}
The position of the horizons and the angular velocity on the horizon are given by
\begin{equation}
r_\pm = M \pm \sqrt{M^2-a^2}, ~~~ \Omega_H = \frac{a}{2Mr_+}.\
\end{equation}
The Hawking temperature and the Bekenstein-Hawking entropy are given by
\begin{equation}
T_H = \frac{r_+-r_-}{8\pi Mr_+},~~S_{\text{BH}} =2\pi Mr_+.\
\end{equation}

Following \cite{Berdberg}, the near-horizon, near-extreme limit is defined by taking the limit of $T_H\to0$, $r\to r_+$ with the near-horizon temperature $T_R \equiv \frac{2MT_H}{\lambda}$ fixed when $\lambda\to0$. Unlike extreme Kerr, even though at asymptotically flat infinity the temperature goes to zero, the temperature near the horizon is still finite because of the infinite blue shift. With this limit, we can expand the black hole parameters as
\begin{equation}\label{nNHEKBHparameter}
    r_+ = M + 2M\lambda\pi T_R+\mathcal{O}(\lambda^2),~~~~ a = M - 2M(\lambda\pi T_R)^2 + \mathcal{O}(\lambda^3),
\end{equation}
and define a new co-rotating coordinates
\begin{equation}
    t = \lambda\frac{\hat{t}}{2M},~~~ r=\frac{\hat{r}-r_+}{\lambda r_+},~~~ \phi = \hat{\phi} - \frac{\hat{t}}{2M}.
\end{equation}
Taking $\lambda\to0$ while keeping $T_R$ fixed, the near-NHEK metric is
\begin{equation}\label{nNHEKmetric}
    ds^2 = 2J\Gamma(\theta)\left(-r(r+4\pi T_R)dt^2 + \frac{dr^2}{r(r+4\pi T_R)} + d\theta^2 + \Lambda^2(\theta)(d\phi + (r+2\pi T_R)dt)^2\right),
\end{equation}
where
\begin{equation}
    \Gamma(\theta) = \frac{1+\cos^2\theta}{2}, ~~~~ \Lambda(\theta) = \frac{2\sin\theta}{1+\cos^2\theta}.
\end{equation}
Taking $T_R$ exactly zero recovers the near-horizon extreme Kerr geometry (NHEK). This near-horizon region of Kerr is not asymptotically flat. It has been shown to be dual to a two-dimensional CFT with central charge $c_R=c_L=12J$ \cite{Guica2009,Matsuo:2009sj,Castro:2009jf,Haco:2018ske}. This duality is supported by the facts that, at and near extremality, the Cardy formula reproduces the Bekenstein Hawking entropy and the finite temperature CFT correlators precisely match the Kerr scattering amplitudes \cite{Berdberg}.

The nNHEK geometry is the NHEK analog of BTZ black holes \cite{Berdberg}, and most similar to thermal Rindler coordinates that only cover a section of AdS \cite{Bilotta}. Unlike NHEK, it has two timelike boundaries that can be joined with Minkowski space to recover the full Kerr geometry. This can be seen in the global extension of the coordinates, which can be described in Kruskal coordinates. The Kruskal coordinates of Kerr are defined as follows
\begin{equation}
    \hat{U}\hat{V} = -\frac{1}{\kappa_+^2}e^{2\kappa_+r^*}, ~~~~ -\frac{\hat{U}}{\hat{V}} = e^{-2\kappa_+\hat{t}},
\end{equation}
where $\kappa_+=2\pi T_H$ is the surface gravity in $r_+$ and $r^*$ is the tortoise coordinate
\begin{equation}
    r^* = \hat{r} + \frac{r_++r_-}{r_+-r_-}\left(r_+\ln\left(\frac{\hat{r}-r_+}{r_+}\right)-r_-\ln\left(\frac{\hat{r}-r_-}{r_-}\right)\right).
\end{equation}
In the near-horizon, near-extreme limit, we have (\ref{nNHEKBHparameter}), $U=\kappa_+\hat{U}$ and $V=\kappa_+\hat{V}$, so the Kruskal coordinates for nNHEK are \cite{Bilotta}
\begin{equation}\label{kruskalkerrnex}
    UV = -\frac{r}{r+4\pi T_R}, ~~~~ -\frac{U}{V} = e^{-4\pi T_Rt},
\end{equation}
where the horizon is now at $U=0$ and $V=0$, and the timelike boundaries are at $UV=-1$. The metric in this coordinate is then
\begin{equation}
    ds^2 = 2J\Gamma(\theta)\left(-\frac{4dUdV}{(1+UV)^2} + d\theta^2 + \Lambda^2(\theta)\left(d\phi + \frac{1}{1+UV}(VdU-UdV)\right)^2\right).
\end{equation}
The Kruskal and Penrose diagrams of this geometry are shown in fig.\eqref{fig1:wormholediagram}.
\begin{figure}
    \begin{tikzpicture}[scale=.9]
\clip (-6,-4) rectangle (20,4);
\path[use as bounding box] (-6,-4) rectangle (20,4);

\begin{scope}[shift={(-2,0)}]

\fill[blue!30]
(-2,2)--(0,4)--(2,2)--(0,0)--cycle;
\fill[blue!30]
(-2,6)--(0,8)--(2,6)--(0,4)--cycle;
\fill[blue!30]
(-2,-6)--(0,-8)--(2,-6)--(0,-4)--cycle;

\fill[blue!30]
(-2,2)
.. controls (-0.5,4)
.. (-2,6)
-- (0,4)
-- cycle;

\fill[blue!30]
(2,2)
.. controls (0.5,4)
.. (2,6)
-- (0,4)
-- cycle;

\fill[blue!30]
(-2,-2)
.. controls (-0.5,-4)
.. (-2,-6)
-- (0,-4)
-- cycle;

\fill[blue!30]
(2,-2)
.. controls (0.5,-4)
.. (2,-6)
-- (0,-4)
-- cycle;

\fill[blue!30]
(-2,-2)--(0,0)--(2,-2)--(0,-4)--cycle;

\fill[yellow!30]
(-2,2)
.. controls (-0.5,0)
.. (-2,-2)
-- (0,0)
-- cycle;
\fill[yellow!30]
(2,2)
.. controls (0.5,0)
.. (2,-2)
-- (0,0)
-- cycle;

\draw (-4,0)--(-2,2)--(0,0)--(-2,-2)--cycle;
\draw (4,0)--(2,2)--(0,0)--(2,-2)--cycle;

\draw[dashed]
(-2,2)--(0,4)--(2,2)--(0,0)--cycle;

\draw[dashed]
(-2,-2)--(0,0)--(2,-2)--(0,-4)--cycle;

\draw[line width=0.8pt]
(-2,2)
.. controls (-.5,0)
.. (-2,-2);

\draw[line width=0.8pt]
(2,2)
.. controls (.5,0)
.. (2,-2);

\draw[line width=0.8pt]
(-2,6)
.. controls (-.5,4)
.. (-2,2);

\draw[line width=0.8pt]
(2,2)
.. controls (.5,4)
.. (2,6);

\draw[line width=0.8pt]
(-2,-6)
.. controls (-.5,-4)
.. (-2,-2);

\draw[line width=0.8pt]
(2,-2)
.. controls (.5,-4)
.. (2,-6);

\draw[decorate,
decoration={zigzag,segment length=3pt}]
(-2,4)--(-2,2);
\draw[decorate,
decoration={zigzag,segment length=3pt}]
(-2,-2)--(-2,-4);
\draw[decorate,
decoration={zigzag,segment length=3pt}]
(2,4)--(2,2);
\draw[decorate,
decoration={zigzag,segment length=3pt}]
(2,-2)--(2,-4);

\node at (-2,0)
{I};

\node at (2,0)
{II};


\end{scope}




\begin{scope}[shift={(7,0)}]

\fill[blue!30] (-2,4) rectangle (2,0);
\fill[blue!30] (-2,-4) rectangle (2,0);

\fill[yellow!30]
(-2,2)--(-2,-2)--(0,0)--cycle;

\fill[yellow!30]
(2,2)--(2,-2)--(0,0)--cycle;

\draw[black,thick]
(-2,4)--(-2,-4);

\draw[black,thick]
(2,4)--(2,-4);

\draw[dashed]
(-2,2)--(0,4)--(2,2)--(0,0)--cycle;

\draw[dashed]
(-2,-2)--(0,0)--(2,-2)--(0,-4)--cycle;

\node[rotate=45] at (0.6-2,-1+.1) {\small \textcolor{black}{$U=0$}};
\node[rotate=-45] at (-0.6+2,-1+.1) {\small \textcolor{black}{$V=0$}};

\node[black,rotate=90] at (-2.3,0)
{$r=r_B$};

\node[black,rotate=90] at (2.3,0)
{$r=r_B$};

\draw[->, very thick, black] (0,.3) -- (.85,1.15);
\draw[->, very thick, black] (0,.3) -- (-.85,1.15);
\node at (-1.05,1.25) {\small \textcolor{black}{$U$}};
\node at (1.05,1.25) {\small \textcolor{black}{$V$}};

\end{scope}

\end{tikzpicture}
    \caption{Near-horizon region of near-extremal Kerr and its global extension in Kruskal coordinates.}
    \label{fig1:wormholediagram}
\end{figure}
From the infinite disconnected boundaries present in the maximal extension, we only consider two boundaries from region I and II in this work. The two boundaries are causally disconnected, a potential for nNHEK to be used as a setup to construct a traversable wormhole.

In BTZ, or any eternal AdS black holes, the AdS/CFT relates the geometry to two copies of CFT that lives in the two asymptotic boundaries, described in a thermofield double state (TFD) \cite{Maldacena:2001kr}. The analogue between nNHEK and BTZ, specifically in the existence of two asymptotic boundaries which each dual to two-dimensional CFT, suggest that similar boundary quantum state is exist. It allows us to define the full systems in the tensor product of two Hilbert spaces $\mathcal{H}=\mathcal{H}_1\otimes\mathcal{H}_2$. We then proposed the TFD in this setup as
\begin{equation}\label{tfd}
    |\text{TFD}\rangle=\frac{1}{\sqrt{Z(T_H,\Omega_H)}}\sum_ne^{-(E_{n}+\Omega_HJ_n)/2T_H}|E_n,J_n\rangle_1\otimes|E_n,J_n\rangle_2,
\end{equation}
where $Z=\Tr{e^{-(H+\Omega_HJ)/T_H}}$ is the partition function, $H$ and $J$ are the Hamiltonian and the angular momentum with eigenvalues $E_n$ and $J_n$ respectively on each CFT. The nNHEK geometry allows right-moving excitation with finite temperature $T_R$, not just left-moving modes.\footnote{The index $L,R$ denotes left- and right-moving CFT, whereas index $1,2$ denotes CFT from left boundary and right boundary.} So, CFT on each boundary can be written in terms of these two different chirality \cite{Caputa:2013eka,Jahnke:2019gxr}.

\subsection{Dirac Equation in nNHEK}
Using the Newman-Penrose formalism, the Dirac equation in nNHEK is separable between the radial and angular function \cite{Chandrasekhar}. For fermions $\psi$ with mass $\mu$, the Dirac equation is
\begin{equation}
    (\gamma^\mu D_\mu - \mu)\psi = 0,
\end{equation}
where $D_\mu = \partial_\mu + \Gamma_\mu$, $\Gamma_\mu=\frac{1}{4}\omega_{(ab)\mu}\gamma^a\gamma^b$, and $\omega_{(ab)\mu}$ is the spin connection. The same goes for the conjugate spinor $\bar{\psi}=\psi^\dagger\gamma^0$
\begin{equation}
(D_\mu\bar{\psi})\gamma^\mu  + \mu\bar{\psi} = 0, ~~~~ D_\mu\bar{\psi} = \partial_\mu\bar{\psi} - \bar{\psi}\Gamma_\mu.
\end{equation}
The Greek letter indices denote bulk space-time indices, while the Latin letter indices denote bulk tangent indices. The analytical solution to the Dirac equation in Kerr with near-horizon, near-extreme limit was obtained in \cite{Hartman:2009nz}. In this limit, the wave function can be decomposed into
\begin{equation}
    \psi(t,r,\theta,\phi) = e^{-i\tilde{\omega}t}e^{im\phi}\left(
    \begin{matrix}
        -R_{\frac{1}{2}}S_{\frac{1}{2}} \\
        \frac{R_{-\frac{1}{2}}S_{-\frac{1}{2}}}{\sqrt{2}M(i-\cos\theta)} \\
        -\frac{R_{-\frac{1}{2}}S_{\frac{1}{2}}}{\sqrt{2}M(i+\cos\theta)} \\
        R_{\frac{1}{2}}S_{-\frac{1}{2}}
    \end{matrix}
    \right),
\end{equation}
where $\tilde{\omega} = \frac{1}{\lambda}(2M\omega - m)$, $R_{\pm\frac{1}{2}}=R_{\pm\frac{1}{2}}(r)$, and $S_{\pm\frac{1}{2}}=S_{\pm\frac{1}{2}}(\theta)$. Near extreme limit, we should fix $\tilde{\omega}$ by taking a limit where only modes near the "superradiant" bound $\omega\simeq m\Omega_H$ survive near the horizon. This is because near the extreme limit $T_H\to0$, the Boltzmann factor $\sim e^{\frac{\omega-m\Omega_H}{T_H}}$ only allows modes with momentum near superradiant bound. Although fermion does not experience superradiance, we still call this limit as "superradiant" for convenience.

The angular function satisfies (for $s=\pm\frac{1}{2}$)
\begin{equation}
    \left[\mathcal{L}_{-s}\frac{1}{\Lambda-2sa\mu\cos\theta}\mathcal{L}_s+\Lambda+2sa\mu\cos\theta\right]S_s(\theta)=0,
\end{equation}
where $\Lambda$ is the separation constant and
\begin{equation}\label{angulareq}
    \mathcal{L}_s \equiv \partial_\theta + 2s(m\csc\theta-a\omega\sin\theta)+\frac{1}{2}\cot\theta.
\end{equation}
Without angular momentum $J=0$ (Schwarzschild), the solution is the standard spin-weighted spherical harmonics \cite{NP1966, Goldberg1967}
\begin{equation}
    e^{im\phi}S_{\pm\frac{1}{2}}(\theta) = Y_{jm,\pm\frac{1}{2}}, ~~~~ \Lambda^2 = \bigg(j+\frac{1}{2}\bigg)^2.
\end{equation}
In Kerr geometry ($J\not=0$), the angular equation can be solved by treating $a\omega$ and $a\mu$ as perturbative parameters \cite{Chakrabarti, Dolan:2009kj}. Writing
\begin{equation}
    S_{jm,\pm\frac{1}{2}}(\theta,\phi) = e^{im\phi}S_{\pm\frac{1}{2}}(\theta), ~~~~ \Lambda_{jm}=\Lambda,
\end{equation}
the wave function and eigenvalue (separation constant) can be expressed as a series of spin-weighted spherical harmonics. However, an exact analytical solution is not possible because the infinite series is not in compact form, except when $a\mu=a\omega$ \cite{Chakrabarti, Dolan:2009kj}.

The radial function satisfies 
\begin{equation}\label{radialeq}
    \left[r(r+4\pi T_R)\frac{d^2}{dr^2} + (1+s)(2r+4\pi T_R)\frac{d}{dr} + V_s\right]R_s(r) = 0,
\end{equation}
where
\begin{equation}
    V_s \equiv \frac{(mr+\tilde{\omega})^2 - is(2r+4\pi T_R)(mr+\tilde{\omega})}{r(r+4\pi T_R)} + s(2s+1) + 2ism - \mu^2r_+^2-\Lambda_{jm}^2.
\end{equation}
The solution of this equation is a hypergeometric function, where the solution for the ingoing boundary conditions is \cite{Hartman:2009nz}
\begin{align}
    R_s(r) = &N_s r^{-i\frac{\tilde{\omega}}{4\pi T_R}-s}\left(\frac{r}{4\pi T_R}+1\right)^{-s+i\left(\frac{\tilde{\omega}}{4\pi T_R}-m\right)} \nonumber\\
    &\times {}_2F_1\left(\frac{1}{2}+\beta-s-im, \frac{1}{2}-\beta-s-im, 1-s-i\frac{\tilde{\omega}}{2\pi T_R}, -\frac{r}{4\pi T_R}\right),
\end{align}
where $\beta^2 = \Lambda_{jm}^2 + \mu^2r_+^2 - m^2$ and the normalization constant satisfies
\begin{equation}\label{normconst}
    \frac{N_{\frac{1}{2}}}{N_{-\frac{1}{2}}} = \frac{\frac{1}{2}-i\frac{\tilde{\omega}}{2\pi T_R}}{M(\Lambda_{jm}+i\mu M)}.
\end{equation}
In this paper, we will only consider real $\beta$ values because the imaginary $\beta$ case will need more care \cite{Berdberg}. Near the horizon ($\hat r\to r_+$ or $r\to 0$), the solution becomes the following
\begin{equation}\label{horizr}
    R_s(r) \sim N_s  r^{-i\frac{\tilde{\omega}}{4\pi T_R}-s},
\end{equation}
while the asymptotic behavior of the solution is (large but finite $r$)
\begin{equation}\label{asympr}
    R_s(r) \sim A_s r^{-s-\frac{1}{2}+\beta} + B_s r^{-s-\frac{1}{2}-\beta},
\end{equation}
where the coefficients are defined in terms of gamma function
\begin{align}
    &A_s = N_s\frac{\Gamma(1-s-i\frac{\tilde{\omega}}{2\pi T_R})\Gamma(2\beta)}{\Gamma(\frac{1}{2}+\beta-i(\frac{\tilde{\omega}}{2\pi T_R}-m))\Gamma(\frac{1}{2}+\beta-s-im)}(4\pi T_R)^{-i\frac{\tilde{\omega}}{4\pi T_R}+\frac{1}{2}-\beta},\\
    &B_s = N_s\frac{\Gamma(1-s-i\frac{\tilde{\omega}}{2\pi T_R})\Gamma(-2\beta)}{\Gamma(\frac{1}{2}-\beta-i(\frac{\tilde{\omega}}{2\pi T_R}-m))\Gamma(\frac{1}{2}-\beta-s-im)}(4\pi T_R)^{-i\frac{\tilde{\omega}}{4\pi T_R}+\frac{1}{2}+\beta}.
\end{align}
We can see that the relations between the expansion coefficients are
\begin{equation}\label{hubABpm}
    A_{\frac{1}{2}}=\frac{\beta-im}{M(\Lambda_{lm}+i\mu M)}A_{-\frac{1}{2}},~~~~ B_{\frac{1}{2}}=\frac{-\beta-im}{M(\Lambda_{lm}+i\mu M)}B_{-\frac{1}{2}}.
\end{equation}
The asymptotic behavior is important when imposing the boundary condition later. We will see that only half of the spinor can be fixed at the nNHEK boundary, and the relation (\ref{hubABpm}) will relate the other half to the fixed one. Another solution satisfies the Neumann boundary condition at the nNHEK boundary \cite{Porfyriadis2014,Gralla2015}
\begin{align}
    R_s^N(r) = & r^{-\frac{1}{2}-\beta-s}\left(\frac{4\pi T_R}{r}+1\right)^{-s+i\left(\frac{\tilde{\omega}}{4\pi T_R}-m\right)} \nonumber\\
    &\times {}_2F_1\left(\frac{1}{2}+\beta-s-im, \frac{1}{2}+\beta-s+i\left(\frac{\tilde{\omega}}{2\pi T_R}-m\right), 1+2\beta, -\frac{4\pi T_R}{r}\right),
\end{align}
where the asymptotic behavior has only falloff term
\begin{equation}
    R_s^N(r)\sim r^{-\frac{1}{2}-\beta-s}.
\end{equation}

\section{Fermionic Double-Trace Deformation in Kerr}\label{sec:fermionicdt}
GJW in their paper considered a quantum mechanical effect to construct traversable wormholes. They introduced a nonlocal deformation of the boundary theory that connects two asymptotic boundaries \cite{GaoJW}
\begin{equation}
    \delta S = \int dtd^{d-1}xh(t,x)\mathcal{O}_L(-t,x)\mathcal{O}_R(t,x),
\end{equation}
where $h(t,x)$ is the coupling between the two boundaries and $\mathcal{O}_{R,L}$ is a scalar operator that acts on the right/left boundary theory. By choosing a certain $h(t,x)$, the ANEC can be violated and the wormhole is traversable. The deformation can also involve other types of operator, such as vector operators, specifically conserved current operators \cite{Ahn}, or tensor operators \cite{khairunnisa2025}. So far, most works only considered deformations involving bosonic operators. In this section, we will define the deformation from fermionic operators in Kerr geometry.

\subsection{Boundary Terms}
The standard Dirac action vanishes for the solution of the equation of motion. This implies that the relation between the partition function of the CFT and the bulk gravitational theory would not hold \cite{HENNINGSON199863}. To avoid this, we can add a boundary term that is non-vanishing for the solution to the equation of motion, so all the contribution to the partition function comes from this term. Consider an action of fermions with mass $\mu$ in nNHEK geometry as \cite{Becker:2012vda}
\begin{equation}\label{action}
    S = i\int d^4x\sqrt{-g}\bar{\psi}(\gamma^\mu D_\mu-\mu)\psi-i\int_{r=r_B}d^3x\sqrt{-g_B}\bar{\psi}\gamma^r\psi,
\end{equation}
where $r_B$ is large but finite cutoff boundary of the nNHEK geometry and $\sqrt{-g_B}$ is the square root of the determinant of the induced metric on that boundary. The first term is the bulk action, while the second term is the boundary term. By adding this boundary term, the variational principle is also satisfied. 

Unlike scalar or vector cases, the Dirac equation is a first order differential equation. As a consequence, we cannot simultaneously fix all the components of $\psi$ and $\bar{\psi}$. Only half of the component can be fixed at the boundary, and the other half is fixed by a relation. To see which component we will fix, we define projection operators \cite{Becker:2012vda}
\begin{equation}
    P_\pm = \frac{1}{2}(1\pm\gamma^0\gamma^3),
\end{equation}
which satisfy $P_+^2=P_+$, $P_-^2=P_-$, and $\frac{1}{2}(\gamma^0\pm\gamma^3)=\gamma^0P_\pm=P_\mp\gamma^0$. With these operators, we can write the spinor as $\psi = \psi_++\psi_-$ where
\begin{align}
    &\psi_+=P_+\psi=e^{-i\tilde{\omega}t}R_{\frac{1}{2}}a_+,\label{psiplus}\\
    &\psi_-=P_-\psi=e^{-i\tilde{\omega}t}R_{-\frac{1}{2}}Za_-,\label{psimin}
\end{align}
and
\begin{equation}
    a_+=\left(
    \begin{matrix}
        -S_{lm,\frac{1}{2}} \\
        0 \\
        0 \\
        S_{lm,-\frac{1}{2}}
    \end{matrix}
    \right), ~~~~ a_-=\gamma^0a_+,~~~~ 
    Z=\left(
    \begin{matrix}
        A & 0 & 0 & 0\\
        0 & \frac{1}{\sqrt{2}M(i-\cos\theta)} & 0 & 0\\
        0 & 0 & \frac{1}{\sqrt{2}M(i+\cos\theta)} & 0 \\
        0 & 0 & 0 & B
    \end{matrix}
    \right),
\end{equation}
arbitrary non-zero $A,B$ is added so that $Z$ is invertible. The projection operators also work for the conjugate spinor $\bar{\psi}P_\pm = \bar{\psi}_\mp$. In terms of projection operators, the gamma matrix $\gamma^r$ in the nNHEK geometry can be written as (see Appendix \ref{NPform})
\begin{equation}
    \gamma^r=-\sqrt{2}\frac{r(r+4\pi T_R)}{4J\Gamma(\theta)}P_-\gamma^0P_+ + \sqrt{2}P_+\gamma^0P_-.
\end{equation}
Thus, the boundary action in terms of $\psi_\pm$ is
\begin{equation}
    S_{bdy}=-i\sqrt{2}\int_{r=r_B}d^3x\sqrt{-g_B}\left(-\frac{r_B^2}{4J\Gamma(\theta)}\bar{\psi}_+\gamma^0\psi_+ + \bar{\psi}_-\gamma^0\psi_-\right).
\end{equation}
Notice that the boundary term is similar to the boundary term of non-relativistic CFT \cite{Laia2011AHF}. 

We can see that from equation (\ref{psiplus}), (\ref{psimin}), and (\ref{asympr}), for real $\beta>0$ the component $\psi_-$ is the leading term. Therefore, we can fix it at the boundary, while $\psi_+$ is related to $\psi_-$ by the Dirac equation and vanishes at the boundary. By fixing $\psi_-$ and $\bar{\psi}_-$ , the second term in the boundary action is a contact term that can be ignored. However, the relation of normalization constant \eqref{normconst} actually indicates that the leading terms of $R_{1/2}$ and $R_{-1/2}$ have the same magnitude near the boundary, represented in \eqref{hubABpm}. So, either of them can be fixed and chosen to be the source \cite{Hartman:2009nz}. The alternative boundary condition where we fix $\psi_+$ rather than $\psi_-$ corresponds to boundary theories with different dimension of boundary operator. In that case, the second term of the boundary action is considered instead the first term. 

\subsection{Double-Trace Deformation}
When we evaluate the action at the boundary, the bulk action vanishes since the equation of motion is satisfied, and the only non-zero contribution comes from the boundary terms. Using the Kerr/CFT correspondence, we will compute the one-point function of the operator $\mathcal{O}$ from the boundary terms. However, since we do not have the Euclidean near-NHEK metric, we need to do the calculation in Lorentzian signature instead of doing analytical continuation. This leads to some additional subtleties and constraints such as; we only consider the incoming boundary condition and we should not consider any contribution from the horizon in our boundary action \cite{Dam,Donald2005,Iqbal2009}. 

To take the functional derivative of the action, we need to express the spinors in momentum space. For $\psi_-$ as the source, the action at the boundary can be written as
\begin{align}
    S &= i\sqrt2\int_{r=r_B} d\theta d\phi dt\sqrt{-g_B}\frac{r_B^2}{4J\Gamma(\theta)}\int d\tilde{\omega} \int d\tilde{\omega}'  \bar{\psi}_+(r,\theta,\phi,\tilde{\omega})\gamma^0\psi_+(r,\theta,\phi,\tilde{\omega}') e^{i(\tilde{\omega}-\tilde{\omega}')t} \nonumber\\
    &= i\sqrt2\int_{r=r_B} d\theta d\phi d\tilde{\omega} \sqrt{-g_B}\frac{r_B^2}{4J\Gamma(\theta)} a_+^\dagger(\theta,\phi)\gamma^0\gamma^0a_+(\theta,\phi) R_{\frac{1}{2}}^*(r,\tilde{\omega})R_{\frac{1}{2}}(r,\tilde{\omega}) \nonumber\\
    &=i\sqrt2\int d\theta d\phi d\tilde{\omega} \sqrt{-g_B} \frac{1}{4J\Gamma(\theta)}a_-^\dagger(\theta,\phi)a_-(\theta,\phi)\nonumber\\
    &\qquad\qquad \times (A_{\frac{1}{2}}^*A_{\frac{1}{2}}r_B^{2\beta}+A_{\frac{1}{2}}^*B_{\frac{1}{2}}+B_{\frac{1}{2}}^*A_{\frac{1}{2}}+B_{\frac{1}{2}}^*B_{\frac{1}{2}}r_B^{-2\beta}).
\end{align}
Using (\ref{hubABpm}), we can express the action in terms of the leading term
\begin{align}
   S &=i\sqrt2\int d\theta d\phi d\tilde{\omega} \sqrt{-g_B}\frac{1}{4J\Gamma(\theta)}a_-^\dagger(\theta,\phi)a_-(\theta,\phi)\nonumber\\
    &\qquad\quad \times (\frac{1}{M^2}A_{-\frac{1}{2}}^*A_{-\frac{1}{2}}r_B^{2\beta} + \frac{\beta+im}{M(\Lambda_{lm}-i\mu M)}A_{-\frac{1}{2}}^*B_{\frac{1}{2}} + \frac{\beta-im}{M(\Lambda_{lm}+i\mu M)}B_{\frac{1}{2}}^*A_{-\frac{1}{2}}\nonumber\\
    &\qquad\quad\quad\quad +B_{\frac{1}{2}}^*B_{\frac{1}{2}}r_B^{-2\beta}).
\end{align}
We should treat terms with $A_{-\frac{1}{2}}$ and $A_{-\frac{1}{2}}^*$ as the source $\chi_-$, and impose the boundary condition
\begin{equation}
    \chi_-(\theta,\phi,\tilde{\omega})=A_{-\frac{1}{2}}(\tilde{\omega})Z(\theta)a_-(\theta,\phi), ~~~~ \lim_{r\to r_B}r^{-\beta}\psi_-=\chi_-.
\end{equation}
It also implies that the first term of the action is a contact term. The one-point function of single-trace operator $\mathcal{\bar{O}}_+$ with $\chi_-$ as the source is
\begin{equation}
    \langle \mathcal{\bar{O}}_+ \rangle = -\frac{\delta S}{\delta \chi_-} = -\frac{r_B\sqrt{2J}\sin\theta}{\sqrt{1+\cos^2\theta}}\frac{\beta-im}{M(\Lambda_{jm}+i\mu M)}a_-^\dagger B_{\frac{1}{2}}^*Z^{-1}.
\end{equation}
Similarly, for $\bar{\chi}$ as the source, we get
\begin{equation}
    \langle \mathcal{O}_+ \rangle = -\frac{\delta S}{\delta \bar{\chi}_-} = - \frac{r_B\sqrt{2J}\sin\theta}{\sqrt{1+\cos^2\theta}} \frac{\beta+im}{M(\Lambda_{jm}-i\mu M)}\gamma^0Z^{*-1}B_{\frac{1}{2}}a_-.
\end{equation}

We define deformation of the boundary theory
\begin{align}
    W &= \int d\theta d\phi dt (\bar{\chi}_-\mathcal{O}_+ + \bar{\mathcal{O}}_+\chi_-),
\end{align}
where $\chi_- \sim A_{-\frac{1}{2}}$ is the source and $\mathcal{O}_+ \sim B_{\frac{1}{2}}$ is the response. Following \cite{GaoJW}, a traversable wormhole with double-trace deformation is constructed by coupling left (labeled with $(1)$) and right (labeled with $(2)$) boundary regions with
\begin{equation}
    \chi_-^{(1)}=h(t)\mathcal\gamma^0{O}_+^{(2)}, ~~~~ \chi_-^{(2)}=h(t)\gamma^0\mathcal{O}_+^{(1)},
\end{equation}
where $h(t)$ is a real function that couples left and right boundaries. \footnote{$\gamma^0$ factor is added to ensure that our double-trace deformation coupling is not zero.} Here we choose it to be a function of $t$ to control when the deformation is activated. In general, it can also depend on spatial coordinates so that the deformation is localized \cite{khairunnisa2025}. Finally, we have a double-trace deformation as
\begin{align}\label{dtdeform}
    \delta H(t) &= -\int d\theta d\phi h(t)(\bar{\mathcal{O}}_+^{(1)}(-t,\theta,\phi)\gamma^0\mathcal{O}_+^{(2)}(t,\theta,\phi)+\bar{\mathcal{O}}_+^{(2)}(t,\theta,\phi)\gamma^0\mathcal{O}_+^{(1)}(-t,\theta,\phi))\nonumber\\
    &= -4Jr_B^{4+2\beta}h(t)\int d\theta d\phi \sin^2\theta (\bar{\eta}_+^{(1)}(-t,\theta,\phi)\gamma^0\eta_+^{(2)}(t,\theta,\phi) + \bar{\eta}_+^{(2)}(t,\theta,\phi)\gamma^0\eta_+^{(1)}(-t,\theta,\phi)),
\end{align}
where
\begin{equation}
    \eta_+(\theta,\phi,\tilde{\omega}) \equiv B_{\frac{1}{2}}(\tilde{\omega})a_+(\theta,\phi)r_B^{-1-\beta}.
\end{equation}
The fields in this double-trace deformation are averaged over a spheroid, similar to what \cite{Bilotta} considered for scalar operators. We will see in the next section that this also appears in the opening of the wormholes, which is related to the average stress energy over a spheroid.

If we choose alternative boundary condition, we fix $\psi_+$ by imposing 
\begin{equation}
    \chi_+(\theta,\phi,\tilde{\omega})=A_{\frac{1}{2}}(\tilde{\omega})a_+(\theta,\phi), \quad \lim_{r\to r_B}r^{1-\beta}\psi_+=\chi_+.
\end{equation}
With this boundary condition, the source is now $\chi_+\sim A_{\frac{1}{2}}$ and the response is $\mathcal{O}_-\sim B_{-\frac{1}{2}}$. Same as before, we couples left and right boundary as
\begin{equation}
    \chi_+^{(1)}=h(t)\gamma^0\mathcal{O}_-^{(2)},\quad \chi_+^{(2)}=h(t)\gamma^0\mathcal{O}_-^{(1)},
\end{equation}
Thus, we have another double-trace deformation as
\begin{equation}\label{dtdeform2}
    \delta H(t) = -8J^3r_B^{2+2\beta}h(t)\int d\theta d\phi \sin^2\theta (\bar{\eta}_-^{(1)}(-t,\theta,\phi)\gamma^0\eta^{(2)}_-(t,\theta,\phi) + \bar{\eta}^{(2)}_-(t,\theta,\phi)\gamma^0\eta^{(1)}_-(-t,\theta,\phi)),
\end{equation}
where
\begin{equation}
    \eta_-(\theta,\phi,\tilde{\omega}) \equiv B_{-\frac{1}{2}}(\tilde{\omega})Z(\theta)a_-(\theta,\phi)r_B^{-\beta}.
\end{equation}
We can always write $\eta_-$ in terms of $\eta_+$,
\begin{equation}
    \eta_- = Z\gamma^0\frac{M(\Lambda_{lm}+i\mu M)}{-\beta-im}r_B\eta_+,
\end{equation}
so that the difference between \eqref{dtdeform} and \eqref{dtdeform2} is their coefficient which can be normalized at the end. Therefore, for the rest of this work we will carry out the computation in $\eta_+$.

\section{The Wormholes}\label{sec:wh}
The backreaction of the deformation (\ref{dtdeform}) introduces negative energy into the bulk. Near the horizon $U=0$ and fixed $(\theta,\phi)$, a perturbation in the bulk can cause a shift to a null ray trajectory
\begin{equation}
    \Delta U = -\frac{1}{2g_{UV}(U=0)}\int dVh_{VV},
\end{equation}
where $h_{\mu\nu}$ is the metric perturbation due to the stress energy tensor. With the linearized Einstein equation, we can relate the perturbation to the average null energy (ANE) as \cite{Bilotta} \footnote{Slightly different from \cite{Bilotta}, in this work we calculate $\Delta U$ instead of $\Delta\hat{U}$, so the order of Hawking temperature is $T_H$ instead of $T_H^2$ from the relation $U=\kappa_+\hat{U}$.} 
\begin{equation}\label{deltaU}
    T_H\Delta U_{\text{avg}} = \frac{1}{8\pi M}8\pi G_\text{N}\int d\Omega dV T_{VV}.
\end{equation}
From this relation, we can see that a negative ANE can trigger a signal from one boundary to cross the wormhole and reach the other boundary, as described by $\Delta U<0$ (see fig.\eqref{fig1:wormhole}). 
\begin{figure}
    \begin{tikzpicture}[scale=.9]
\clip (-6,-4) rectangle (20,4);
\path[use as bounding box] (-6,-4) rectangle (20,4);

\begin{scope}[shift={(3,0)}]

\fill[blue!30] (-2,4) rectangle (2,0);
\fill[blue!30] (-2,-4) rectangle (2,0);

\fill[yellow!30]
(-2,2)--(-2,-2)--(0,0)--cycle;

\fill[yellow!30]
(2,2)--(2,-2)--(0,0)--cycle;

\draw[black,thick]
(-2,4)--(-2,-4);

\draw[black,thick]
(2,4)--(2,-4);

\draw[dashed]
(-2,2)--(0,4)--(2,2)--(0,0)--cycle;

\draw[dashed]
(-2,-2)--(0,0)--(2,-2)--(0,-4)--cycle;

\node[rotate=45] at (0.5-2,-1+.1) {\small \textcolor{black}{signal}};
\draw[
    postaction={decorate},
    decoration={
        markings,
        mark=at position 0.8 with {\arrow{>}} 
    },
    very thick, magenta
](-2,0.3-2) -- (.85,1.15);

\draw[
    postaction={decorate},
    decoration={
        markings,
        mark=at position 0.8 with {\arrow{>}} 
    },
    very thick, magenta
] (1.15,.85) -- (2,1.7);
\draw[->, very thick, magenta] (.85,1.15) -- (1.15,.85);

\draw[very thick, decorate, decoration={snake, amplitude=0.05cm, segment length=0.3cm}, teal!60!black] (2,0) -- (1.15,.85);
\draw[very thick, decorate, decoration={snake, amplitude=0.05cm, segment length=0.3cm}, teal!60!black] (.85,1.15) -- (0,2);
\draw[very thick, decorate, decoration={snake, amplitude=0.05cm, segment length=0.3cm}, teal!60!black] (-2,0) -- (0,2);

\node[black,left] at (-2,0)
{$\mathcal{O}^{(1)}_+$};

\node[black,right] at (2,0)
{$\mathcal{O}^{(2)}_+$};

\node[black,rotate=90] at (-2.3,0.3-2)
{$r=r_B$};

\node[black,rotate=90] at (2.3,0.3-2)
{$r=r_B$};

\end{scope}

\end{tikzpicture}
    \caption{A signal (red) moves along the horizon from the left boundary to the right boundary through the wormhole. The shift in $U$ direction is caused by the double-trace deformation (blue).}
    \label{fig1:wormhole}
\end{figure}
The coupling between left and right boundary operators can be interpreted as a shock wave with negative energy that shifts the signal trajectory in the $U$ direction. This negative contribution only comes from a quantum mechanical effect, which is the double-trace deformation. As we mentioned earlier, the stress energy tensor is averaged over a spheroid. This is because we are only interested in the traversability and do not need the full angular dependence.

\subsection{Stress Energy Tensor}
To see the effect of double-trace deformation, consider the stress energy tensor of Dirac fermions as \cite{Freedman_VanProeyen_2012}
\begin{equation}
    T_{\mu\nu} = \frac{i}2(\bar{\psi}\gamma_{(\mu} D_{\nu)}\psi - (D_{(\mu}\bar{\psi })\gamma_{\nu)}\psi ),
\end{equation}
where parentheses in the indices denote symmetrization. Using the point splitting method, the one-loop expectation value is
\begin{align}
    \langle T_{\mu\nu}(x,x')\rangle= \frac{i}{2}\lim_{x'\to x}  (&\partial_{(\nu}\langle\bar{\psi}(x')\gamma_{\mu')}\psi(x)\rangle + \langle \bar{\psi}(x')\gamma_{(\mu'}\Gamma_{\nu)}\psi(x)\rangle \nonumber\\
    &- \partial_{(\mu'}\langle\bar{\psi}(x')\gamma_{\nu)}\psi(x)\rangle + \langle \bar{\psi}(x')\Gamma_{(\mu'}\gamma_{\nu)}\psi(x)\rangle),
\end{align}
where we only need the $\langle T_{VV}\rangle$ component to calculate the opening of the wormhole
\begin{align}
    \langle T_{VV}(x,x')\rangle= \frac{i}{2}\lim_{x'\to x} (&\partial_{V}\langle\bar{\psi}(x') \gamma_{V'}\psi(x)\rangle + \langle\bar{\psi}(x') \gamma_{V'}\Gamma_{V} \psi(x)\rangle \nonumber\\
    &- \partial_{V'}\langle\bar{\psi}(x') \gamma_{V} \psi(x)\rangle + \langle\bar{\psi}(x') \Gamma_{V'}\gamma_{V}\ \psi(x)\rangle  ).
\end{align}
At the horizon $U=0$, in terms of projection operators, $\gamma_V$ and $\Gamma_V$ can be written as (see Appendix \ref{NPform})
\begin{equation}\label{gammakruskalhorizon}
    \gamma_V = -\sqrt{2}P_-\gamma^0P_+, ~~~~ \Gamma_V = \frac{\Lambda(\theta)}{4\sqrt{J\Gamma(\theta)}} \gamma^2P_-\gamma^0P_+.
\end{equation}
Substituting (\ref{gammakruskalhorizon}), the expectation value of the stress energy tensor on the horizon becomes
\begin{equation}
    \langle T_{VV}(x,x')\rangle = -\frac{i\sqrt{2}}{2} \lim_{x'\to x}  (\partial_{V}\langle\psi^\dagger_+(x')\psi_+(x)\rangle -\partial_{V'}\langle\psi^\dagger_+(x')\psi_+(x)\rangle),
\end{equation}
where the spin connection terms cancel each other because at the horizon $U=0$, $\Gamma_V\gamma_V=-\gamma_V\Gamma_V$. In terms of fermion two-point function $G(x,x')=\langle\psi(x)\bar{\psi}(x')\rangle$, we can write it as
\begin{equation}\label{tvv}
    \langle T_{VV}(x,x')\rangle = \frac{i}{2} \lim_{x'\to x}  \text{Tr}(\gamma_{V'}\partial_{V}G(x,x') -\partial_{V'}G(x,x')\gamma_{V}).
\end{equation}

To calculate (\ref{tvv}) and the two-point function, we need implicitly to choose a vacuum state. For bosons in Kerr spacetime, this becomes a challenge because we cannot construct a Hartle-Hawking-like state due to bosonic superradiance \cite{KAY199149}. Although various vacuum states that have the desired properties have been proposed such as \cite{Candelas1981, Frolov1989}, it has been shown that these states \cite{ottewill2000} are irregular in the off-axis region or are not equilibrium states with the expected desired symmetries of a Hartle-Hawking state. Fortunately, fermion does not experience superradiant amplification and a Hartle-Hawking vacuum state that is regular even in the off-axis region has been proposed \cite{Casals2013}. Thus, we can safely construct the wormhole throughout the nNHEK spacetime with TFD as the boundary state \eqref{tfd}.

\subsection{Modified two-point function}
The deformation $\delta H$ (\ref{dtdeform}) acts as a perturbation that modifies the two-point function. Since the stress–energy tensor is obtained by projecting and tracing the two-point function, it suffices to evaluate the correlator $\langle\psi_+(x)\psi^\dagger_+(x')\rangle$. A unitary operator $U(t,t_0) = \mathcal{T}e^{-i\int_{t_0}^tdt_1\delta H(t_1)}$ evolves the field operators in the interaction picture
\begin{equation}
    G_{++}(x,x') = \langle U^{-1}(t,t_0)\psi_+(x)U(t,t_0)U^{-1}(t',t_0)\psi^\dagger_+(x')U(t',t_0)\rangle.
\end{equation}
By expanding the operator $U$, the first order correction to the two-point function of $\psi_+$ is
\begin{equation}
    G_{++}^{(1)}(x,x') = i\int_{t_0}^t dt_1 \langle\psi_+(x)[\delta H(t_1),\psi^\dagger_+(x')] \rangle + i\int_{t_0}^{t'} dt_1 \langle [\delta H(t_1),\psi_+(x)]\psi^\dagger_+(x') \rangle.
\end{equation}
We need to be more careful for our next step because fermions satisfy anti-commutation relation instead of commutation relation. Substituting the deformation $\delta H$ \eqref{dtdeform}, we get
\begin{align}
    G_{++}^{(1)}(x,x') &= -i4Jr_B^{4+2\beta}\int_{t_0}^t dt_1d\mathbf{x}_1 \sin\theta_1 h(t_1) \langle\psi_+(x) [\eta^{(1)\dagger}_+(-t_1,\mathbf{x}_1) \eta^{(2)}_+(t_1,\mathbf{x}_1), \psi_+^\dagger(x')]\rangle \nonumber\\
    &\qquad\qquad\qquad\qquad +\langle \psi_+(x) [\eta^{(2)\dagger}_+(t_1,\mathbf{x}_1) \eta^{(1)}_+(-t_1,\mathbf{x}_1), \psi_+^\dagger(x')]\rangle \nonumber\\
    &\quad-i4Jr_B^{4+2\beta}\int_{t_0}^{t'} dt_1d\mathbf{x}_1 \sin\theta_1 h(t_1) \langle[\eta^{(1)\dagger}_+(-t_1,\mathbf{x}_1) \eta^{(2)}_+(t_1,\mathbf{x}_1), \psi_+(x)] \psi_+^\dagger(x')\rangle \nonumber\\
    &\qquad\qquad\qquad\qquad + \langle[\eta^{(2)\dagger}_+(t_1,\mathbf{x}_1) \eta^{(1)}_+(-t_1,\mathbf{x}_1), \psi_+(x)] \psi_+^\dagger(x')\rangle,
\end{align}
where $\mathbf{x}=(\theta,\phi)$ and $d\mathbf{x}=\sin\theta d\theta d\phi$. Then, we can use $[AB,C]=A\{B,C\}-\{A,C\}B$ to write it with anti-commutation relation as
\begin{align}
    G_{++}^{(1)}(x,x') &= -i4Jr_B^{4+2\beta}\int_{t_0}^t dt_1d\mathbf{x}_1 \sin\theta_1 h(t_1)\langle \psi_+(x)\eta^{(1)\dagger}_+(-t_1,\mathbf{x}_1)\{\eta^{(2)}_+(t_1,\mathbf{x}_1),\psi_+^\dagger(x')\} \nonumber\\
    &\qquad\qquad\qquad\qquad - \psi_+(x)\{\eta^{(1)\dagger}_+ (-t_1,\mathbf{x}_1),\psi_+^\dagger(x')\}\eta^{(2)}_+(t_1,\mathbf{x}_1) \nonumber\\
    &\qquad\qquad\qquad\qquad + \psi_+(x)\eta^{(2)\dagger}_+(t_1,\mathbf{x}_1)\{\eta^{(1)}_+(-t_1,\mathbf{x}_1),\psi_+^\dagger(x')\} \nonumber\\
    &\qquad\qquad\qquad\qquad - \psi_+(x)\{\eta^{(2)\dagger}_+(t_1,\mathbf{x}_1), \psi_+^\dagger(x')\}\eta^{(1)}_+(-t_1,\mathbf{x}_1)\rangle \nonumber\\
    &\quad-i4Jr_B^{4+2\beta}\int_{t_0}^{t'} dt_1d\mathbf{x}_1 \sin\theta_1 h(t_1)\langle \eta^{(1)\dagger}_+(-t_1,\mathbf{x}_1)\{\eta^{(2)}_+(t_1,\mathbf{x}_1),\psi_+(x)\}\psi_+^\dagger(x') \nonumber\\
    &\qquad\qquad\qquad\qquad - \{\eta^{(1)\dagger}_+(-t_1,\mathbf{x}_1), \psi_+(x)\} \eta^{(2)}_+(t_1,\mathbf{x}_1))\psi_+^\dagger(x') \nonumber\\
    &\qquad\qquad\qquad\qquad +\eta^{(2)\dagger}_+(t_1,\mathbf{x}_1)\{\eta^{(1)}_+(-t_1,\mathbf{x}_1),\psi_+(x)\}\psi_+^\dagger(x') \nonumber\\
    &\qquad\qquad\qquad\qquad - \{\eta^{(2)\dagger}_+(t_1,\mathbf{x}_1),\psi_+(x)\} \eta^{(1)}_+(-t_1,\mathbf{x}_1))\psi_+^\dagger(x')\rangle.
\end{align}
Using the fact that $\{\eta^{(1)}_+,\psi^\dagger_+\}=0$ to ensure causality and $\{\eta_+,\psi_+\}=\{\eta^\dagger_+,\psi_+^\dagger\}=0$ for all fermions since $\eta_+=\psi_+(r_B)$, only the first term of first integral and the last term of second integral remain
\begin{align}
     G_{++}^{(1)}(x,x') &= -i4Jr_B^{4+2\beta}\int_{t_0}^t dt_1d\mathbf{x}_1 \sin\theta_1 h(t_1)\langle \psi_+(x)\eta^{(1)\dagger}_+(-t_1,\mathbf{x}_1)\rangle\langle\{\eta^{(2)}_+(t_1,\mathbf{x}_1),\psi_+^\dagger(x')\}\rangle \nonumber\\
     &~ +i4Jr_B^{4+2\beta}\int_{t_0}^{t'} dt_1d\mathbf{x}_1 \sin\theta_1 h(t_1) \langle\{\psi_+(x),\eta^{(2)\dagger}_+(t_1,\mathbf{x}_1)\}\rangle\langle \eta^{(1)}_+(-t_1,\mathbf{x}_1))\psi_+^\dagger(x')\rangle,
\end{align}
where we consider single-trace operator that dual to the bulk field, so that in large $N$ limit we used large $N$ factorization to split the expectation value\footnote{Example of Kerr/CFT and non-relativistic CFTs in large $N$ limit is studied in \cite{El-Showk:2011euy}.}. In this work, we choose a simple form of coupling $h(t)$ as
\begin{equation}\label{couplingh}
    h(t) = h\delta(2\pi T_R(t-t_0)),
\end{equation}
where $h$ is constant. This coupling means that the source is a delta function, so the deformation is activated immediately at $t_0$ and the two-point function is modified as
\begin{align}\label{1stordercorrection}
     G_{++}^{(1)}(x,x') &= -i\frac{2Jr_B^{4+2\beta}h}{\pi T_R}\int d\mathbf{x}_1 \sin\theta_1 (\langle \psi_+(x)\eta^{(1)\dagger}_+(-t_0,\mathbf{x}_1)\rangle\langle\{\eta^{(2)}_+(t_0,\mathbf{x}_1),\psi_+^\dagger(x')\}\rangle \nonumber\\
     &\qquad\qquad\qquad\qquad- \langle\{\psi_+(x),\eta^{(2)\dagger}_+(t_0,\mathbf{x}_1)\}\rangle\langle \eta^{(1)}_+(-t_0,\mathbf{x}_1))\psi_+^\dagger(x')\rangle).
\end{align}
Since the wave equation is separable, the two-point function can be decomposed into
\begin{equation}
    G_{++}^{(1)}(x,x') = \sum_{j,m} \tilde{G}_{++}^{(1)}(t,r,t',r')a_+(\mathbf{x}) a_+^\dagger(\mathbf{x}'),
\end{equation}
where
\begin{align}
    \tilde{G}_{++}^{(1)}(t,r,t',r') = -i\frac{2Jr_B^{4+2\beta}h}{\pi T_R}C_{jm} (&K_W(t,r,-t_0-\frac{i}{2T_R})K_\text{ret}^{\dagger}(t',r',t_0) \nonumber\\
    &- K_\text{ret}(t,r,t_0)K_W^{\dagger}(t',r',-t_0-\frac{i}{2T_R})),
\end{align}
and
\begin{equation}
    C_{jm} \equiv \int d\mathbf{x}_1 \sin\theta_1(|S_{jm,\frac{1}{2}}(\mathbf{x}_1)|^2+|S_{jm,-\frac{1}{2}}(\mathbf{x_1})|^2).
\end{equation}
We defined retarded bulk-to-boundary propagator and Wightman function as
\begin{align}
    &K_\text{ret}(r,t,t_0) \equiv -i\langle \{\psi_+(t,r),\eta^{(2)\dagger}_+(t_0)\} \rangle,\\
    &K_W(r,t,t_0) \equiv -i\langle \psi_+(t,r)\eta^{(2)\dagger}_+(t_0) \rangle,
\end{align}
and since the TFD on the left and the right region are separated by imaginary time shift $\Delta \hat{t}=i/2T_H$ (or $\Delta t=i/2T_R$ in the nNHEK coordinates), we used Kubo-Martin-Schwinger condition to relate $\eta^{(2)}_+$ and $\eta^{(1)}_+$. To have the full expression of the modified two-point function, we need to compute these propagators first.

\subsubsection{Bulk-to-boundary Propagator}
The bulk-to-boundary propagator can be obtained by taking one point of the bulk-to-bulk propagator to the boundary. The radial bulk-to-bulk propagator or the radial Green function for $s=\frac{1}{2}$ satisfies an inhomogeneous equation
\begin{equation}
    (r(r+4\pi T_R))^{-\frac{1}{2}}\frac{d}{dr}\left[(r(r+4\pi T_R))^{\frac{3}{2}}\frac{d}{dr}G_{\frac{1}{2}}(r,r',\tilde{\omega})\right] + V_{\frac{1}{2}}G_{\frac{1}{2}}(r,r',\tilde{\omega}) = \delta(r-r').
\end{equation}
The Green function is constructed by two homogeneous solutions of equation (\ref{radialeq}). The retarded solution that satisfies purely in-going at the horizon and the Neumann boundary condition at the nNHEK boundary is
\begin{equation}
    G_{\frac{1}{2}}(r,r',\tilde{\omega})=-\frac{R_{\frac{1}{2}}^{in}(r_<)R_{\frac{1}{2}}^N(r_>)}{(r(r+4\pi T_R))^{-\frac{1}{2}}W},
\end{equation}
where $r_<$ is min$(r,r')$, $r_>$ is max$(r,r')$, and $W$ is the $r$-independent Wronskian $W=-2\beta A_{\frac{1}{2}}$. Taking $r$ near the horizon and $r'$ at the nNHEK boundary, we get the retarded bulk-to-boundary propagator as
\begin{equation}
    K_\text{ret}(r,\tilde{\omega})=\frac{N_{\frac{1}{2}}r^{-\frac{i\tilde{\omega}}{4\pi T_R}}r_B^{-1-\beta}}{(4\pi T_R)^{-\frac{1}{2}}2\beta A_{\frac{1}{2}}}.
\end{equation}
We then Fourier transform it to express it in real space as
\begin{equation}
    K_\text{ret}(r,t,t_0) = \int \frac{d\tilde{\omega}}{2\pi} e^{-i\tilde{\omega}(t-t_0)}\frac{N_{\frac{1}{2}}r^{-\frac{i\tilde{\omega}}{4\pi T_R}}r_B^{-1-\beta}}{(4\pi T_R)^{-\frac{1}{2}}2\beta A_{\frac{1}{2}}}.
\end{equation}

In Kruskal coordinates (\ref{kruskalkerrnex}) where the horizon is $U=0$ and the nNHEK boundary is at $U_0V_0=-1$, the retarded bulk-to-boundary propagator becomes
\begin{equation}
    K_\text{ret}(V,V_0) = \frac{(4\pi T_R)^{\beta}r_B^{-1-\beta}}{2\beta} \frac{\Gamma(\beta-im)}{\Gamma(2\beta)} \int \frac{d\tilde{\omega}}{2\pi} \left(\frac{V_0}{V}\right)^{i\frac{\tilde{\omega}}{2\pi T_R}}  \frac{\Gamma(\frac{1}{2}+\beta-i(\frac{\tilde{\omega}}{2\pi T_R}-m))}{\Gamma(\frac{1}{2}-i\frac{\tilde{\omega}}{2\pi T_R})},
\end{equation}
where we already substitute $A_{\frac{1}{2}}$. To evaluate the integral, define a new variable $z \equiv \frac{1}{2}+\beta-i(\frac{\tilde{\omega}}{2\pi T_R}-m)$, and write the integral as
\begin{align}
    K_\text{ret}(V,V_0) = &i\frac{(4\pi T_R)^{\beta+1}r_B^{-1-\beta}}{8\pi\beta} \frac{\Gamma(\beta-im)}{\Gamma(2\beta)}\left(\frac{V_0}{V}\right)^{\frac{1}{2}+\beta+im}\nonumber\\
    &\times \int_{\frac{1}{2}+\beta+i\infty}^{\frac{1}{2}+\beta-i\infty} dz \left(\frac{V_0}{V}\right)^{-z}\frac{\Gamma(z)}{\Gamma(z-\beta-im)}.
\end{align}
The gamma function has poles at $z=-n$, where $n=0,1,2,3,...$, and the residue is
\begin{equation}
    \text{Res}(f(z),z=-n) = \frac{(-1)^n}{n!} \left(\frac{V_0}{V}\right)^{n}\frac{1}{\Gamma(-n-\beta-im)}.
\end{equation}
Integrating by using the residue theorem, we finally have the retarded bulk-to-boundary propagator
\begin{align}\label{retk}
    K_\text{ret}(V,V_0) &= -\frac{(4\pi T_R)^{\beta+1}r_B^{-1-\beta}}{4\beta} \frac{\Gamma(\beta-im)}{\Gamma(2\beta)} \left(\frac{V_0}{V}\right)^{\frac{1}{2}+\beta+im}\nonumber\\
    &~~~~\times\sum_n^\infty\frac{(-1)^n}{n!} \left(\frac{V_0}{V}\right)^{n}\frac{1}{\Gamma(-n-\beta-im)}\Theta(V-V_0)\nonumber\\
    &=-\frac{(4\pi T_R)^{\beta+1}r_B^{-1-\beta}}{4\beta} \frac{\Gamma(\beta-im)}{\Gamma(2\beta)\Gamma(-\beta-im)} \sqrt{\frac{V}{V_0}} \left(\frac{V}{V_0}-1\right)^{-\beta-im-1}\Theta(V-V_0),
\end{align}
where $\Theta$ is the Heaviside step function to close the contour. It is interesting to see that the factor $(V/V_0-1)$ emerges naturally from the infinite series term. This factor is related to the dependence of the stress energy tensor on the insertion time $t_0$. In the vector case \cite{Ahn} and tensor case \cite{khairunnisa2025}, only the factor $V/V_1$ appears in the propagators because the late time modes are dominant $V/V_1\gg1$. As a consequence, the stress energy tensor is independent of the insertion time, and then it is necessary to apply some prescription to include the dependence on $t_0$.

\subsubsection{Wightman Function}
After we obtained the retarded bulk-to-boundary propagator, the Wightman function can be easily obtained from the relation below
\begin{equation}
    K_W(r,\tilde{\omega}) = i\frac{e^{\tilde{\omega}/T_R}}{e^{\tilde{\omega}/T_R}+1}\Im{K_\text{ret}(r,\tilde{\omega})}.
\end{equation}
Using this relation and the same procedure of integration as before, the Wightman function is as follows
\begin{equation}
    K_W(V,V_0) = i\frac{(4\pi T_R)^{\beta+1}r_B^{-1-\beta}}{16\beta} \frac{\Gamma(\beta-im)}{\Gamma(2\beta)\Gamma(-\beta-im)} \frac{(V/V_0)^{\frac{1}{2}}}{\sin(\pi(\beta+im))} \left(1-\frac{V}{V_0}\right)^{-\beta-im-1}.
\end{equation}
To relate the right boundary and the left boundary, we see that for $V_0=e^{2\pi T_R t_0}$ at the boundary
\begin{equation}
    K_W(V,V_0) = i\frac{(4\pi T_R)^{\beta+1}r_B^{-1-\beta}}{16\beta} \frac{\Gamma(\beta-im)}{\Gamma(2\beta)\Gamma(-\beta-im)} \frac{\sqrt{V}e^{-\pi T_R t_0}}{\sin(\pi(\beta+im))} \left(1-Ve^{-2\pi T_R t_0}\right)^{-\beta-im-1},
\end{equation}
we can use the KMS condition $t_0\to -t_0-\frac{i}{2T_R}$ so the final expression of our Wightman function is
\begin{equation}\label{wfunc}
    K_W(V,V_0) = -\frac{(4\pi T_R)^{\beta+1}r_B^{-1-\beta}}{16\beta} \frac{\Gamma(\beta-im)}{\Gamma(2\beta)\Gamma(-\beta-im)} \frac{\sqrt{VV_0}}{\sin(\pi(\beta+im))} \left(1+VV_0\right)^{-\beta-im-1}.
\end{equation}

With (\ref{retk}) and (\ref{wfunc}), the first order correction to the two-point function (\ref{1stordercorrection}) becomes
\begin{equation}
    \tilde{G}_{++}^{(1)}(V,V') = -i\frac{Jhr_B^2}{8\beta^2} (4\pi T_R)^{2\beta+1}C_{jm}(F(V,V')-F^\dagger(V',V)),
\end{equation}
where
\begin{equation}\label{fvv}
    F(V,V') \equiv \frac{|\Gamma(\beta-im)|^2}{|\Gamma(2\beta)\Gamma(-\beta-im)|^2} \frac{\sqrt{VV'}(1+VV_0)^{-1-\beta-im}}{\sin(\pi(\beta+im))} \left(\frac{V'}{V_0}-1\right)^{-1-\beta+im},
\end{equation}
and $F^\dagger(V',V)$ is the complex conjugate of $F(V,V')$ but with $V\leftrightarrow V'$.

\subsection{Average Null Energy}
The zeroth order two-point function does not contribute to open the wormhole, the only contribution is coming from the first order correction due to double-trace deformation. With the modified two-point function, equation (\ref{tvv}) becomes
\begin{align}
    \langle T_{VV}\rangle &= -\frac{i\sqrt{2}}{2} \lim_{x'\to x} \text{Tr}(\partial_{V}G^{(1)}_{++}(x,x') - \partial_{V'}G^{(1)}_{++}(x,x'))\nonumber\\
    &= -\frac{Jhr_B^2}{8\sqrt{2}} \sum_{jm}\frac{(4\pi T_R)^{2\beta+1}}{\beta^2}C_{jm}(|S_{jm,\frac{1}{2}}|^2+|S_{jm,-\frac{1}{2}}|^2) ] \nonumber\\
    &\quad\quad \times\lim_{V'\to V} (\partial_{V}(F(V,V') - F^\dagger(V',V)) - \partial_{V'}(F(V,V') - F^\dagger(V',V))).
\end{align}
Substituting (\ref{fvv}), we get the following expectation value of the stress energy tensor
\begin{align}
    \langle T_{VV}\rangle = &\frac{Jhr_B^2}{4\sqrt{2}} \sum_{jm}\frac{(4\pi T_R)^{2\beta+1}}{\beta^2}C_{jm}(|S_{jm,\frac{1}{2}}|^2+|S_{jm,-\frac{1}{2}}|^2) \frac{|\Gamma(\beta-im)|^2}{|\Gamma(2\beta)\Gamma(-\beta-im)|^2}\nonumber\\
    &\times \frac{V}{P(V,V_0)^{2+\beta}}\left[-m\Im{Q(V,V_0)}\bigg(2V+\frac{1}{V_0}-V_0\bigg)\right. \nonumber\\
    &\qquad\qquad\qquad\qquad - \left.(1+\beta)\Re{Q(V,V_0)}\bigg(\frac{1}{V_0}+V_0\bigg)\right],
\end{align}
where
\begin{align}
    P(V,V_0) &\equiv (1+VV_0)\left(\frac{V}{V_0}-1\right),\\
    Q(V,V_0) &\equiv \frac{1}{\sin(\pi(\beta+im))}\left(\frac{V/V_0-1}{1+VV_0}\right)^{im}.
\end{align}
To see wormhole traversability, we need the average null energy. Integrating the stress energy tensor over a spheroid and $V$ we get
\begin{align}\label{ane}
    \int d\Omega dV\langle T_{VV}\rangle = & \mathcal{N}\sum_{jm}\frac{(4\pi T_R)^{2\beta+1}}{4\sqrt{2}\beta^2}C_{jm}\frac{|\Gamma(\beta-im)|^2}{|\Gamma(2\beta)\Gamma(-\beta-im)|^2}  \nonumber\\
    &\times\int_{V_0}^\infty dV\frac{V}{P(V,V_0)^{2+\beta}}\left[-m\Im{Q(V,V_0)}\bigg(2V+\frac{1}{V_0}-V_0\bigg)\right. \nonumber\\
    &\qquad\qquad\qquad\qquad\qquad\quad - \left.(1+\beta)\Re{Q(V,V_0)}\bigg(\frac{1}{V_0}+V_0\bigg)\right],
\end{align}
where we defined $\mathcal{N}=Jhr_B^2$ and we used the normalization of $S_{jm,s}$
\begin{equation}
    \int d\Omega(|S_{jm,\frac{1}{2}}|^2+|S_{jm,-\frac{1}{2}}|^2)=1.
\end{equation}
With this average null energy we can get the wormhole opening from equation (\ref{deltaU}).

\subsubsection{Massless Fermion}
\begin{figure}[htbp]
    \centering
    \includegraphics[scale=0.5]{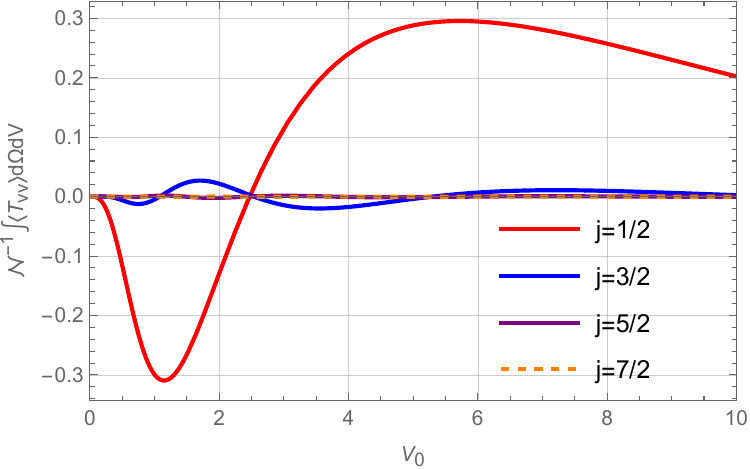}
    \includegraphics[scale=0.5]{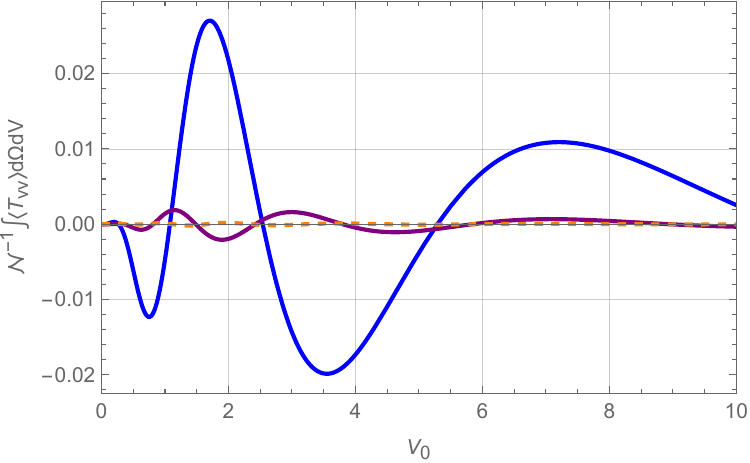}
    \caption{Average null energy of each $j$ with $4\pi T_R=10^{-4}$. Right plot show ANE except for $j=\frac{1}{2}$ to see more clearly that $\text{ANE}_{j+1}\ll \text{ANE}_j$ and they have oscillatory feature.}
    \label{fig:ane}
\end{figure}

When $\mu=0$, the angular equation (\ref{angulareq}) reduces to the equation of spin-weighted spheroidal harmonics. Figure (\ref{fig:ane}) shows the ANE for $j=\frac{1}{2}$, $j=\frac{3}{2}$, $j=\frac{5}{2}$, and $j=\frac{7}{2}$. We can see that the $j=\frac{1}{2}$ mode is dominant and $\text{ANE}_{j+1}\ll \text{ANE}_j$, thus the summation of $j$ in (\ref{ane}) converges. The wormhole is most traversable at early times, peaked around $V_0=1$ or $t_0=0$, similar to bosonic case. An apparent difference is that traversability fluctuates depending on when we turn on the deformation. 

\begin{figure}
    \centering
    \includegraphics[width=0.5\linewidth]{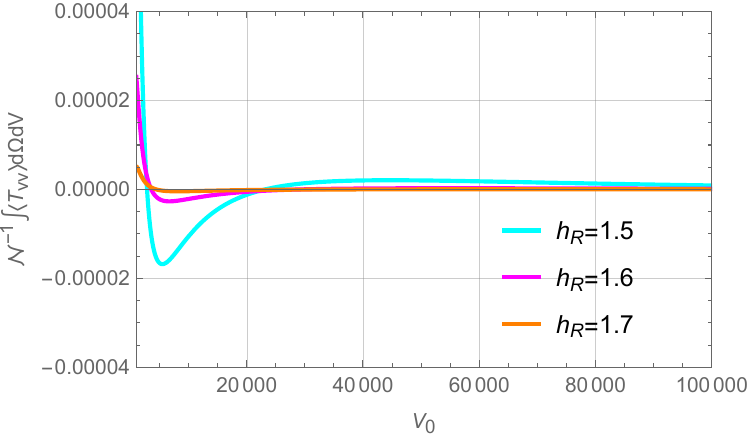}
    \includegraphics[width=0.45\linewidth]{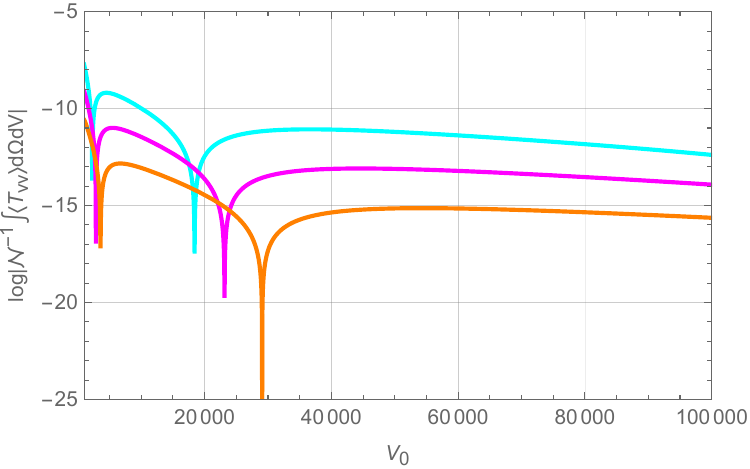}
    \caption{Average null energy at late times from $V_0=10^3$ to $V_0=10^5$ with $4\pi T_R=10^{-4}$.}
    \label{fig:anelate}
\end{figure}

\begin{figure}[htbp]
    \centering
    \includegraphics[scale=0.7]{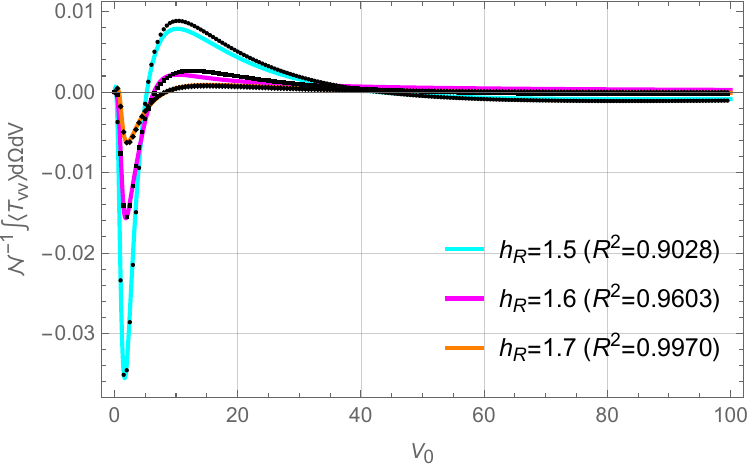}
    \caption{Comparison between numerical result (dots) and fitting model (solid lines) for various conformal weight $h_R$ with $4\pi T_R=10^{-4}$.  From each fitting we get $R^2=0.9028$ for $h_R=1.5$, $R^2=0.9603$ for $h_R=1.6$, and $R^2=0.9970$ for $h_R=1.7.$}
    \label{fig:anefitting}
\end{figure}

Figure (\ref{fig:anelate}) show that the fluctuation is damped and $|\text{ANE}|$ is much smaller at late times. The value of $\beta$, we write it in terms of conformal weight $h_R=\frac{1}{2}+\beta$ \cite{Hartman:2009nz}, contributes to the damping factor.
Moreover, to see late time behavior of ANE, we define fitting model that can describe the numerical results as
\begin{equation}
    \int d\Omega dV\langle T_{VV}\rangle =-a_1\frac{V_0^{a_2}}{V_0^{a_2+1}+a_3}\sin\left(\frac{a_4}{V_0^{a_5}+a_6}-a_7\right).
\end{equation}
This model includes an exponential decay factor as in bosonic case \cite{Ahn, khairunnisa2025}, without a power-law factor because our result is non-diffusive, and with an additional sinusoidal factor to reflect the oscillation. Figure (\ref{fig:anefitting}) also show the comparison between the fitting model and the actual numerical results. From this model, we can see that at late times $V_0\gg1$, the oscillation is suppressed and ANE behave as
\begin{equation}
    \int d\Omega dV\langle T_{VV}\rangle \sim e^{-2\pi T_Rt_0}.
\end{equation}
Therefore, the wormhole is still not traversable if we turn on the deformation too late, $\text{ANE}_{t\to\infty}\to0$.

\begin{figure}[htbp]
    \centering
    \includegraphics[scale=0.5]{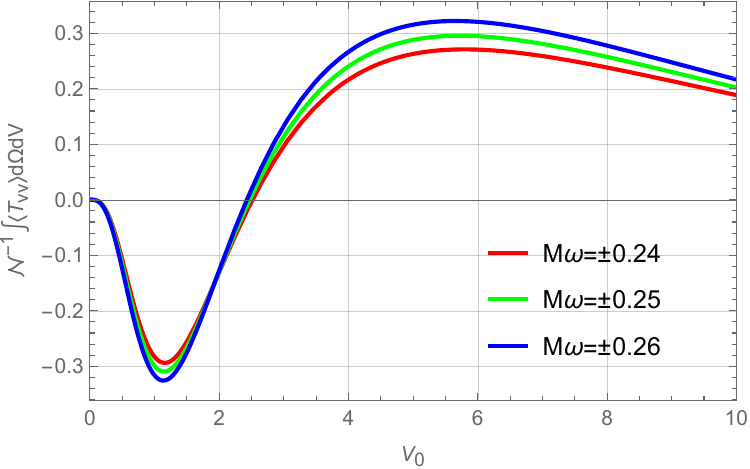}
    \includegraphics[scale=0.5]{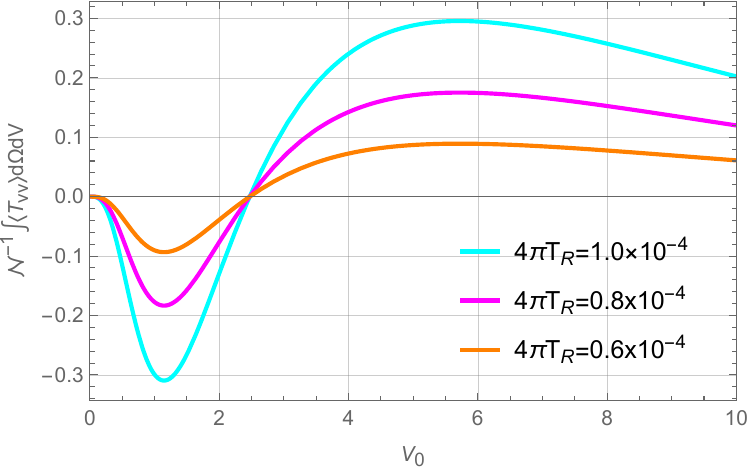}
    \caption{Average null energy for various $M\omega$ near superradiant bound (left) and $4\pi T_R$ (right).}
    \label{fig:anecaw}
\end{figure}

\begin{figure}[htbp]
    \centering
    \includegraphics[scale=0.5]{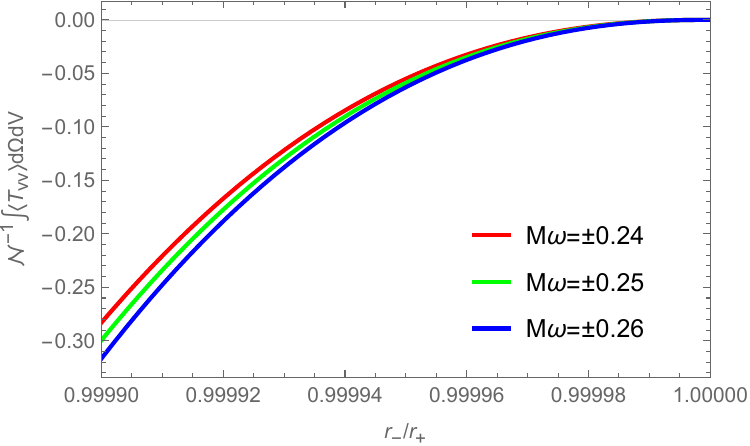}
    \includegraphics[scale=0.5]{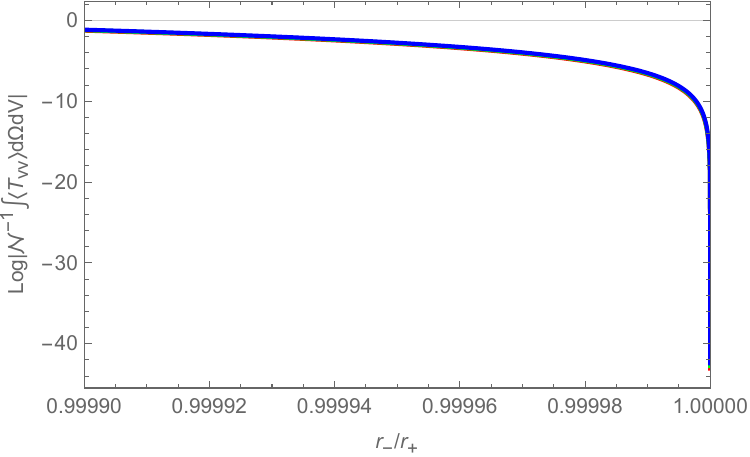}
    \caption{Average null energy at the extreme limit for $V_0=1$.}
    \label{fig:aneextreme}
\end{figure}

We mentioned how conformal weight affects ANE as a damping factor. Meanwhile, the conformal weight itself depends on the separation constant $\Lambda_{jm}$ (which depends on $(\omega,j,m)$) and the fermion mass $\mu$. Furthermore, the ANE also depends on temperature. Figure (\ref{fig:anecaw}) shows ANE for different values of $M\omega$ and $4\pi T_R$. We vary $M\omega$ around the superradiant bound $\omega\simeq m\Omega_H$, which for $m=\pm\frac{1}{2}$ and near extreme limit is $m\Omega_H\approx\pm0.25/M$.
The separation constant is calculated numerically from the spin-weighted spheroidal harmonic equation. We can see that smaller temperature results in less negative ANE or smaller wormhole opening, this can indicate that near extremality the wormhole is less traversable. However, to actually see how ANE behaves at the extreme limit, we need to take $T_R$ to zero instead of fixing it. Figure (\ref{fig:aneextreme}) shows the ANE in the extreme limit. We can see that the wormhole is closed at the extreme limit ($r_-/r_+=1$) since the ANE is zero or $\log|\int dV\langle T_{VV}\rangle|\to -\infty$.

\subsubsection{Massive Fermion}
\begin{figure}[htbp]
    \centering
    \includegraphics[scale=0.6]{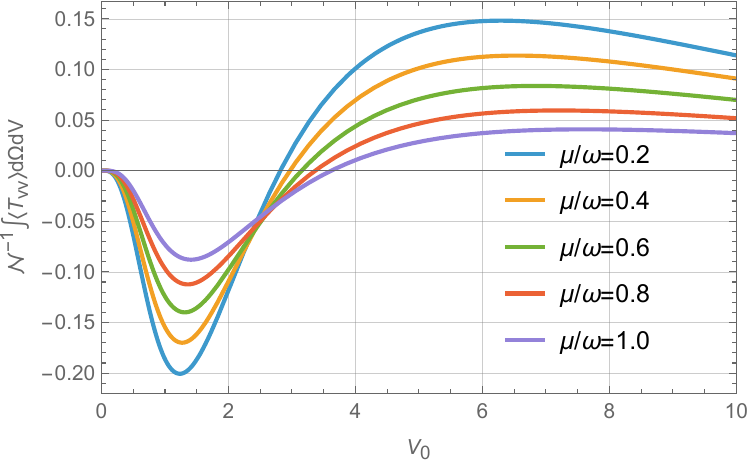}
    \caption{Average null energy for various mass to frequency ratio $\mu/\omega$ with $4\pi T_R=10^{-4}$ and $M\omega=\pm0.2$.}
    \label{fig:anecmu}
\end{figure}

When $\mu\neq0$, the angular wave function and the separation constant are mass dependent. We use mass dependent separation constant calculated numerically from \cite{Chakrabarti, Dolan:2009kj}. Figure (\ref{fig:anecmu}) shows that for fixed $M\omega$ and $4\pi T_R$, fermion with higher mass results in less negative ANE or smaller wormhole opening. The mass explicitly determines $\beta$ and also implicitly controls the separation constant $\Lambda_{jm}$. We can see this result as the mass adds positive energy to the system, it lessens the negative energy from the non-local interaction.

\subsection{Eikonal Method}
This method was first realized in \cite{Maldacena2017} to compute traversability of AdS$_2$ wormhole from a different perspective. They considered a scattering of a signal $\phi$ by a shock wave that is generated from the boundary deformation $\mathcal{O}$. The signal experience a time advance so it can reach the other boundary. Eikonal approximation is used in this method to compute the scattering amplitude. Since then, this method has been used in other cases, such as rotating BTZ \cite{caceres2018} and higher dimension hyperbolic black holes \cite{Ahn:2020csv}. 

The traversablity is directly computed from the expectation value of two-sided correlation function of the form
\begin{equation}
    \langle\{\psi^{(1)}_+(-t_1,\mathbf{x}_1),e^{-i\mathcal{V}}\psi^{(2)}_+(t_2,\mathbf{x}_2)e^{i\mathcal{V}}\}\rangle,
\end{equation}
where $\langle\cdots\rangle$ is taken in the thermofield double state, and
\begin{equation}\label{dteikon}
    \mathcal{V} = \frac{1}{K}\sum_{i=1}^K \int dt'\delta H_i(t'),
\end{equation}
is the double-trace deformation involving $K$ fields. This correlation function takes non-zero value if the signal can travel through the wormholes. In bosonic cases, we can simply compute half of the correlation function above,
\begin{equation}
    C \equiv \langle e^{-i\mathcal{V}}\psi^{(2)}_+(t_2,\mathbf{x}_2)e^{i\mathcal{V}}\psi^{(1)}_+(-t_1,\mathbf{x}_1)\rangle,
\end{equation}
and the original commutator is obtained from the imaginary part of it, $\langle[\psi^{(1)}_+,e^{-i\mathcal{V}}\psi^{(2)}_+e^{i\mathcal{V}}]\rangle=2i\Im{C}$, because the fields are Hermitian $\psi^\dagger=\psi$. Meanwhile in this case, the original anti-commutator cannot be recovered from the real part of $C$,
\begin{equation}
    2\Re{C} = \langle e^{-i\mathcal{V}}\psi^{(2)}_+e^{i\mathcal{V}}\psi^{(1)}_+\rangle + \langle \psi^{(1)\dagger}_+e^{-i\mathcal{V}}\psi^{(2)\dagger}_+ e^{i\mathcal{V}}\rangle \not= \langle\{\psi^{(1)}_+,e^{-i\mathcal{V}}\psi^{(2)}_+e^{i\mathcal{V}}\}\rangle,
\end{equation}
except when $\psi$ is Majorana fermion. So, we also need to compute the other half, defined as $C'\equiv\langle \psi^{(1)}_+e^{-i\mathcal{V}}\psi^{(2)}_+ e^{i\mathcal{V}}\rangle$, to obtain the full anti-commutator.

In large $K$ and small $G_N$ limit, we can simplify our calculation to just
\begin{equation}
    C = e^{-i\langle\mathcal{V}\rangle} \tilde{C},
\end{equation}
where we define $\tilde{C}\equiv\langle \psi^{(2)}_+(t_2,\mathbf{x}_2) e^{i\mathcal{V}} \psi^{(1)}_+(-t_1,\mathbf{x}_1)\rangle$. Let us consider the first order of $\tilde{C}$ in small $h$ first,
\begin{align}
    \tilde{C}_1 = -\frac{i}{K}\sum_{i=1}^K 4Jr_B^{4+2\beta} \int dt'd\mathbf{x}' &\sin\theta' h(t',\mathbf{x}')[\langle \psi^{(2)}_+(t_2,\mathbf{x}_2) \eta^{(1)\dagger}_{+i}(-t',\mathbf{x}')\eta^{(2)}_{+i}(t',\mathbf{x}')\psi^{(1)}_+(-t_1,\mathbf{x}_1) \rangle \nonumber\\
    &+ \langle \psi^{(2)}_+(t_2,\mathbf{x}_2) \eta^{(1)}_{+i}(-t',\mathbf{x}')\eta^{(2)\dagger}_{+i}(t',\mathbf{x}')\psi^{(1)}_+(-t_1,\mathbf{x}_1\rangle].
\end{align}
Each term is an out-of-time order correlation functions in the context of quantum chaos, and can be computed following technique in \cite{Shenker:2014cwa}. We will carry out the computation for a single double-trace deformation and omit the $i$ label for now. We can write the first expectation value as
\begin{equation}
    \langle \text{out}|\text{in}\rangle = \langle \psi^{(2)}_+(t_2,\mathbf{x}_2) \eta^{(1)\dagger}_+(-t',\mathbf{x}')\eta^{(2)}_+(t',\mathbf{x}')\psi^{(1)}_+(-t_1,\mathbf{x}_1) \rangle,
\end{equation}
where "in" and "out" states are defined as
\begin{equation}
    |\text{in}\rangle = \eta^{(2)}_+(t',\mathbf{x}')\psi^{(1)}_+(-t_1,\mathbf{x}_1) |\text{TFD}\rangle, \quad |\text{out}\rangle = \eta^{(1)}_+(-t',\mathbf{x}')\psi^{(2)\dagger}_+(t_2,\mathbf{x}_2) |\text{TFD}\rangle.
\end{equation}
We can then decompose the states in a basis of well-defined momentum in $U$ and $V$ direction and well-defined transverse position $\mathbf{\tilde{x}}$,
\begin{align}
    &|\text{in}\rangle = \int dp_1^V dp_4^U\int d\mathbf{\tilde{x}}_1 d\mathbf{\tilde{x}}_4 \Psi_{\eta^{(2)}_+}(p_4^U,\mathbf{\tilde{x}}_4)\Psi_{\psi_+^{(1)}}(p_1^V,\mathbf{\tilde{x}}_1) (|p_1^V,\mathbf{\tilde{x}}_1\rangle \otimes |p_4^U,\mathbf{\tilde{x}}_4\rangle)_\text{in}, \\
    &|\text{out}\rangle = \int dp_2^V dp_3^U\int d\mathbf{\tilde{x}}_2 d\mathbf{\tilde{x}}_3 \Psi_{\eta^{(1)}_+}(p_3^U,\mathbf{\tilde{x}}_3)\Psi_{\psi_+^{(2)}}(p_2^V,\mathbf{\tilde{x}}_2) (|p_2^V,\mathbf{\tilde{x}}_2\rangle \otimes |p_3^U,\mathbf{\tilde{x}}_3\rangle)_\text{out}.
\end{align}
The wave functions $\Psi_{\eta_+}$ and $\Psi_{\psi_+}$ are given by the Fourier transform of bulk-to-boundary propagators
\begin{align}
    &\Psi_{\psi_+^{(1)}}(p_1^V,\mathbf{\tilde{x}}_1) = \int dU e^{iA_0(\tilde{\theta}_1)p_1^VU} \langle\phi_{\psi_+}^\dagger(U,V,\mathbf{\tilde{x}}_1)\psi^{(1)}_+(-t_1,\mathbf{x}_1)\rangle_{V=0},\\
    &\Psi_{\psi_+^{(2)}}(p_2^V,\mathbf{\tilde{x}}_2) = \int dU e^{iA_0(\tilde{\theta}_2)p_2^VU} \langle\phi_{\psi_+}(U,V,\mathbf{\tilde{x}}_2)\psi^{(2)\dagger}_+(t_2,\mathbf{x}_2)\rangle_{V=0},\\
    &\Psi_{\eta^{(1)}_+}(p_3^U,\mathbf{\tilde{x}}_3) = \int dV e^{iA_0(\tilde{\theta}_3)p_3^UV} \langle\phi_{\eta_+}^\dagger(U,V,\mathbf{\tilde{x}}_3)\eta^{(1)}_+(-t',\mathbf{x}')\rangle_{U=0},\\
    &\Psi_{\eta^{(2)}_+}(p_4^U,\mathbf{\tilde{x}}_4) = \int dV e^{iA_0(\tilde{\theta}_4)p_4^UV} \langle \phi_{\eta_+}^\dagger(U,V,\mathbf{\tilde{x}}_4)\eta^{(2)}_+(t',\mathbf{x}')\rangle_{U=0},
\end{align}
where $A_0(\tilde{\theta})=2J(1+\cos^2\tilde{\theta})$ is the $UV$ component of the metric evaluated at either $U=0$ or $V=0$ and at transverse position $\tilde{\theta}$, and this factor comes from the fact that $U$ and $V$ are conjugate to $p_{U}=-A_0(\tilde{\theta})p^{V}$ and $p_{V} = -A_0(\tilde{\theta})p^U$ respectively. From the bulk-to-boundary propagator definition, $\phi_{\psi_+}$ and $\phi_{\eta_+}$ are the bulk fields dual to $\psi_+$ and $\eta_+$ at the right region.

The basis can be normalized by considering overlap between two Dirac wave functions on a null slice. First, the overlap between $\Psi_+$ and $\Phi_+$ at $U=0$ slice is given by
\begin{equation}
    \langle\Psi_+|\Phi_+\rangle = 2J\int dVd\mathbf{x} n_U\bar{\Psi}_+(U,V,\mathbf{x})\gamma^U\Phi_+(U,V,\mathbf{x})|_{U=0}.
\end{equation}
Then, we Fourier transform the wave functions in momentum space,
\begin{align}\label{norm1}
    \langle\Psi_+|\Phi_+\rangle 
    &=\frac{\sqrt{2}J}{\pi}\int d\mathbf{x}d\mathbf{x}' \delta(\mathbf{x},\mathbf{x}')A_0(\theta)\int dp^U dp'^U \delta(p^U-p'^U) \Psi^\dagger_+(p^U,\mathbf{x})\Phi_+(p'^U,\mathbf{x}'),
\end{align}
where
\begin{equation}
    \delta(\mathbf{x},\mathbf{x}')=\frac{\delta(\theta-\theta')}{\sin\theta'}\delta(\phi-\phi').
\end{equation}
On the other hand, we can also write $\langle\Psi_+|\Phi_+\rangle$ in $|p^U,\mathbf{x}\rangle$ basis as
\begin{equation}\label{norm2}
    \langle\Psi_+|\Phi_+\rangle = \int dp^U dp'^U\int  d\mathbf{x}d\mathbf{x}'\Psi^\dagger_+(p^U,\mathbf{x})\Phi_+(p'^U,\mathbf{x}')\langle p^U,\mathbf{x}|p'^U,\mathbf{x}'\rangle.
\end{equation}
From \eqref{norm1} and \eqref{norm2}, we get the normalization of the basis as
\begin{equation}
    \langle p^U,\mathbf{x}|p'^U,\mathbf{x}'\rangle = \frac{ \sqrt{2}JA_0(\theta)}{\pi}\delta(p^U-p'^U)\delta(\mathbf{x},\mathbf{x}').
\end{equation}

Using the eikonal approximation, the scattering between $\psi_+$ and $\eta_+$ can be approximated by the shock wave amplitude $S_\text{grav}=e^{i\delta}$ \cite{VERLINDE1992246,Kabat:1992tb}. In this approximation, the "in" and "out" are related by
\begin{equation}
    (|p_2^V,\tilde{\mathbf{x}}_2\rangle\otimes|p_3^U,\tilde{\mathbf{x}}_3\rangle)_\text{out} \approx e^{i\delta}(|p_2^V,\tilde{\mathbf{x}}_2\rangle\otimes|p_3^U,\tilde{\mathbf{x}}_3\rangle)_\text{in} + |\text{inelastic}\rangle,
\end{equation}
where $\delta$ is called eikonal phase and $|\text{inelastic}\rangle$ is the inelastic component of the scattering that orthogonal to all "in" states. Based on this approximation and the basis normalization, we get the overlap between "in" and "out" states as
\begin{align}
    \langle\text{out}|\text{in}\rangle = &\left(\frac{\sqrt{2}J}{\pi}\right)^2\int dp_1^Vd\mathbf{\tilde{x}}_1A_0(\tilde{\theta}_1) \Psi^\dagger_{\psi_+^{(2)}}(p_1^V,\mathbf{\tilde{x}}_1)\Psi_{\psi_+^{(1)}}(p_1^V,\mathbf{\tilde{x}}_1)\nonumber\\
    &\times\int dp_4^Ud\mathbf{\tilde{x}}_4 A_0(\tilde{\theta}_4)\Psi^\dagger_{\eta^{(1)}_+}(p_4^U,\mathbf{\tilde{x}}_4)\Psi_{\eta^{(2)}_+}(p_4^U,\mathbf{\tilde{x}}_4)e^{i\delta}.
\end{align}
Thus, we can write $\tilde{C}_1$ as
\begin{align}
    \tilde{C}_1 = &-i4Jr_B^{4+2\beta}\left(\frac{\sqrt{2}J}{\pi}\right)^2\int dp_1^Vd\mathbf{\tilde{x}}_1A_0(\tilde{\theta}_1) \Psi^\dagger_{\psi_+^{(2)}}(p_1^V,\mathbf{\tilde{x}}_1)\Psi_{\psi_+^{(1)}}(p_1^V,\mathbf{\tilde{x}}_1)\nonumber\\
    &\times\int dp_4^Ud\mathbf{\tilde{x}}_4 A_0(\tilde{\theta}_4)\int dt'd\mathbf{x}'\sin\theta 'h(t',\mathbf{x}')\bigg[\Psi^\dagger_{\eta^{(1)}_+}(p_4^U,\mathbf{\tilde{x}}_4)\Psi_{\eta^{(2)}_+}(p_4^U,\mathbf{\tilde{x}}_4) \nonumber\\
    &\qquad\qquad\qquad\qquad\qquad\qquad + \Psi^\dagger_{\eta^{(2)}_+}(p_4^U,\mathbf{\tilde{x}}_4)\Psi_{\eta^{(1)}_+}(p_4^U,\mathbf{\tilde{x}}_4)\bigg]e^{i\delta}.
\end{align}
Furthermore, for all order in $h$,
\begin{equation}
    \tilde{C} = \frac{\sqrt{2}J}{\pi}\int dq^Vd\mathbf{\tilde{x}}_1A_0(\tilde{\theta}_1) \Psi^\dagger_{\psi_+^{(2)}}(q^V,\mathbf{\tilde{x}}_1)\Psi_{\psi_+^{(1)}}(q^V,\mathbf{\tilde{x}}_1)e^{-iD},
\end{equation}
where
\begin{align}
    D\equiv&4Jr_B^{4+2\beta}\frac{\sqrt{2}J}{\pi}\int dp^Ud\mathbf{\tilde{x}}_4 A_0(\tilde{\theta}_4)\nonumber\\
    &\times\int dt'd\mathbf{x}'\sin\theta'h(t',\mathbf{x}')\left[\Psi^\dagger_{\eta^{(1)}_+}(p^U,\mathbf{\tilde{x}}_4)\Psi_{\eta^{(2)}_+}(p^U,\mathbf{\tilde{x}}_4) + \Psi^\dagger_{\eta^{(2)}_+}(p^U,\mathbf{\tilde{x}}_4)\Psi_{\eta^{(1)}_+}(p^U,\mathbf{\tilde{x}}_4)\right]e^{i\delta},
\end{align}
and we replaced $p_1$ and $p_4$ with $q$ and $p$ respectively for convenience.

In point-splitting method, we did not compute the transverse profile to get the ANE's full angular dependence. The traversability is determined by integrating both side of linearized Einstein equation over the spheroid \cite{Bilotta}. This choice implies the stress energy tensor is not localized but still has $\theta$ dependency. It is suffice then to assume the shock wave is homogeneous in the $\phi$ coordinate and let the angular dependency carried by the momentum. The stress energy tensor of the shock wave along $V=0$ takes a simple form of
\begin{equation}\label{swtvv}
    T_{VV} =\frac{p_V}{2J}\delta(V).
\end{equation}
Then from linearized Einstein equation, the backreaction from the shock wave is
\begin{equation}\label{swhvv}
    \frac{\kappa_+}{r_+(1+\cos^2\theta)}h_{VV} = 8\pi G_N T_{VV}.
\end{equation}
On the other hand, we consider signal along $U=0$ that is produced by a local operator at $\phi=\tilde{\phi}_1$,
\begin{equation}\label{signaltuu}
    T_{UU} = \frac{q_U}{2J}\delta(U)\delta(\phi-\tilde{\phi}_1).
\end{equation}
Note that $p_V$ and $q_U$ are the momentum of shock wave and signal at $\tilde{\theta}_4$ and $\tilde{\theta}_1$ respectively. The eikonal phase from the scattering is given by the on-shell action of them,
\begin{equation}
    S_{\text{classical}} = \frac{1}{2}\int d^4x\sqrt{-g}h_{VV}T^{VV},
\end{equation}
where $h_{VV}$ belongs to the shock wave and $T^{VV}=g^{UV}g^{UV}T_{UU}$ belongs to the signal. Using \eqref{swtvv}, \eqref{signaltuu}, and \eqref{swhvv}, we get
\begin{equation}
    \delta = \frac{8\pi G_N r_+}{4J^2\kappa_+} A_0(\tilde{\theta}_1)A_0(\tilde{\theta}_4)p^Uq^V.
\end{equation}

Now consider the probe limit, where the backreaction of the signal is too small to deform the shock wave geometry, or in other words $G_Nq^V\ll1$. The probe approximation implies that $\delta\ll1$, so we can evaluate $D$ order by order as $D=D_0+q^VD_1$ by expanding $e^{i\delta}$ to the first order. The zeroth order term, $D_0=-\langle \mathcal{V\rangle}$, will cancel with $e^{-i\langle\mathcal{V}\rangle}$ factor in $C$. Whereas the first order term correspond to a shift in $U$ direction,
\begin{align}
    C&=e^{-i\langle\mathcal{V}\rangle}\frac{\sqrt{2}J}{\pi}\int dq^Vd\mathbf{\tilde{x}}_1A_0(\tilde{\theta}_1) \Psi^\dagger_{\psi_+^{(2)}}\Psi_{\psi_+^{(1)}}e^{-i(D_0+q^VD_1)}\nonumber\\
    &=e^{-i\langle\mathcal{V}\rangle}\langle\psi^{(2)}_+e^{-i(D_0+q^VD_1)}\psi^{(1)}_+\rangle \nonumber\\
    &=\langle\psi^{(2)}_+e^{i\frac{1}{A_0(\tilde{\theta}_1)} D_1\hat{q}_U}\psi^{(1)}_+\rangle.
\end{align}
We can see that the operator $\hat{q}_U$ shifts $\psi^{(1)}_+$ in $U$ direction by an amount $\frac{1}{A_0(\tilde{\theta}_1)}D_1$.

To compare the result here with our result in point-splitting method, we use the bulk-to-boundary propagators that we already have \eqref{wfunc} for the wave functions,
\begin{align}
    \Psi_{\eta^{(1)}_+}=-\int dV&e^{iA_0(\tilde{\theta}_4)p^UV}a_+(\mathbf{x}')a^\dagger_+(\tilde{\mathbf{x}}_4)\frac{(4\pi T_R)^{\beta+1}r_B^{-1-\beta}}{16\beta}\frac{\Gamma(\beta+im)}{\Gamma(2\beta)\Gamma(-\beta+im)}\nonumber\\
    &\times\frac{\sqrt{VV'}}{\sin(\pi(\beta-im))}(1+VV')^{-\beta+im-1},
\end{align}
\begin{align}
    \Psi_{\eta^{(2)}_+}=\int dV&e^{iA_0(\tilde{\theta}_4)p^UV}a_+(\mathbf{x}')a^\dagger_+(\tilde{\mathbf{x}}_4)\frac{(4\pi T_R)^{\beta+1}r_B^{-1-\beta}}{8\beta}\frac{\Gamma(\beta+im)}{\Gamma(2\beta)\Gamma(-\beta+im)}\nonumber\\
    &\times\frac{\sqrt{V/V'}}{e^{-2i\pi(\beta-im)}-1}\left(\frac{V}{V'}-1\right)^{-\beta+im-1}.
\end{align}
Using them in $D_1$, we have
\begin{align}
    D_1=&-iJr_B^2 \frac{\sqrt{2}J}{\pi}\left(\frac{8\pi G_Nr_+}{4J^2\kappa_+}\right)A_0(\tilde{\theta}_1)\sum_{j,m}\frac{(4\pi T_R)^{2\beta+2}}{32\beta^2} \nonumber\\
    &\times\int dt'd\mathbf{x}'\sin\theta'h(t',\mathbf{x}') a^\dagger_+(\mathbf{x}')a_+(\mathbf{x}')\int d\mathbf{\tilde{x}}_4A_0^2(\tilde{\theta}_4)a_+^\dagger(\mathbf{\tilde{x}}_4)a_+(\mathbf{\tilde{x}}_4)\nonumber\\
     &\times\int dVdV_1\int dp^Up^U \bigg[e^{iA_0(\tilde{\theta}_4)p^U(V_1-V)}\frac{F(V,V_1)}{e^{-2i\pi(\beta-im)}-1}+e^{iA_0(\tilde{\theta}_4)p^U(V-V_1)}\frac{F^\dagger(V,V_1)}{e^{2i\pi(\beta+im)}-1}\bigg].
\end{align}
Using the integral representation of first derivative of delta function and considering an instantaneous perturbation $h(t',\mathbf{x}')=h\delta(t'-t_0)/2\pi T_R$, we have
\begin{align}
    D_1=&Jhr_B^2 \frac{\sqrt{2}J}{\pi}\left(\frac{8\pi G_Nr_+}{4J^2\kappa_+}\right)A_0(\tilde{\theta}_1)\sum_{j,m}\frac{(4\pi T_R)^{2\beta+1}}{16\beta^2}C_{jm}\int d\mathbf{\tilde{x}}_4A_0^2(\tilde{\theta}_4)a_+^\dagger(\mathbf{\tilde{x}}_4)a_+(\mathbf{\tilde{x}}_4)\nonumber\\
    &\times \int dVdV_1\frac{1}{A_0^2(\tilde{\theta}_4)}2\pi\bigg[\delta'(V-V_1)\frac{F(V,V_1)}{e^{-2i\pi(\beta-im)}-1}+\delta'(V_1-V)\frac{F^\dagger(V,V_1)}{e^{2i\pi(\beta+im)}-1}\bigg]\nonumber\\
\end{align}
where $\delta'(x-y)$ is first derivative of delta function with respect to $x$. Working on the integral,
\begin{align}
    D_1=&-Jhr_B^2 \left(\frac{G_N}{ MT_H}\right)A_0(\tilde{\theta}_1)\sum_{j,m}\frac{(4\pi T_R)^{2\beta+1}}{4\sqrt{2}\beta^2}C_{jm}\int d\mathbf{\tilde{x}}_4a_+^\dagger(\mathbf{\tilde{x}}_4)a_+(\mathbf{\tilde{x}}_4)\nonumber\\
    &\times \int dV_1\partial_{V}\left[\frac{F(V,V_1)}{e^{-2i\pi(\beta-im)}-1}\right]_{V=V_1}+\int dV\partial_{V_1}\left[\frac{F^\dagger(V,V_1)}{e^{2i\pi(\beta+im)}-1}\right]_{V_1=V}.
\end{align}

\begin{figure}[htbp]
    \centering
    \includegraphics[scale=0.6]{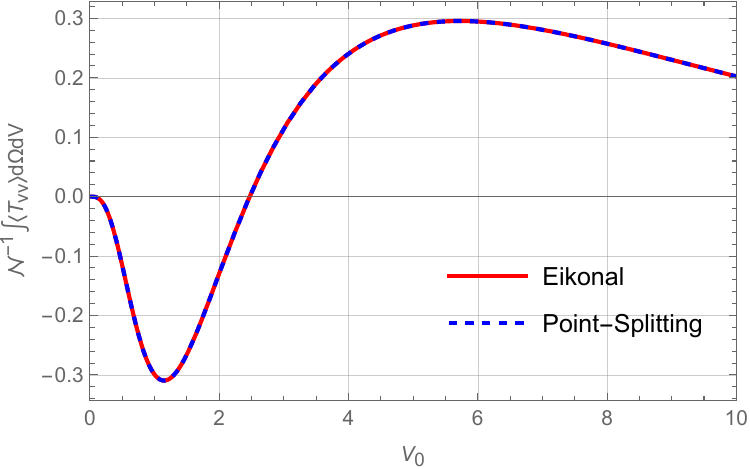}
    \caption{Comparison of average null energy from eikonal method and point-splitting method for massless fermion with $4\pi T_R=10^{-4}$ and $M\omega=\pm 0.25$.}
    \label{fig:aneeikon}
\end{figure}

As we mentioned earlier, $C$ only contains half of the information from the original anti-commutator. Whereas the other half is from $C'$. Using the same method above, we obtain
\begin{equation}
    C'=\langle e^{-i\frac{1}{A_0(\tilde{\theta}_1)}D_1'\hat{q}_U}\psi^{(1)}_+ \psi^{(2)}_+\rangle.
\end{equation}
This term indicates shift of $\psi^{(1)}_+$ in $U$ direction by an amount $-\frac{1}{A_0(\tilde{\theta}_1)}D_1'$, where
\begin{align}
    D_1'=&Jhr_B^2 \left(\frac{G_N}{MT_H}\right)A_0(\tilde{\theta}_1)\sum_{j,m}\frac{(4\pi T_R)^{2\beta+1}}{4\sqrt{2}\beta^2}C_{jm}\int d\mathbf{\tilde{x}}_4a_+^\dagger(\mathbf{\tilde{x}}_4)a_+(\mathbf{\tilde{x}}_4)\nonumber\\
    &\times \int dV_1\partial_{V}\left[\frac{F^\dagger(V,V_1)}{e^{2i\pi(\beta+im)}-1}\right]_{V=V_1}+\int dV\partial_{V_1}\left[\frac{F(V,V_1)}{e^{-2i\pi(\beta-im)}-1}\right]_{V_1=V}.
\end{align} 
Then, the total shift is
\begin{equation}
    \Delta U_{\text{avg}} = \frac{1}{A_0(\tilde{\theta}_1)}(D_1-D_1'),
\end{equation}
and the ANE averaged over spheroid is
\begin{align}\label{aneeikon}
    \int d\Omega dVT_{VV} = & Jhr_B^2 \sum_{j,m}\frac{(4\pi T_R)^{2\beta+1}}{2\sqrt{2}\beta^2}C_{jm}\frac{|\Gamma(\beta-im)|^2}{|\Gamma(2\beta)\Gamma(-\beta-im)|^2}\nonumber\\
    &\times \int_0^\infty dV \frac{V}{P(V,V_0)^{2+\beta}}\bigg[m\Im{\frac{Q(V,V_0)}{e^{-2i\pi(\beta-im)}-1}}\left(\frac{1}{V_0}+V_0\right)\nonumber\\
    &\qquad\qquad\qquad +(1+\beta)\Re{\frac{Q(V,V_0)}{e^{-2i\pi(\beta-im)}-1}}\left(2V+\frac{1}{V_0}-V_0\right)\bigg].
\end{align}
Each $i$ fields in the deformation contributes ANE by this amount. Consequently, the sum in the fields indicates the total ANE is proportional to $K$. However, the $K$ factor is normalized from our definition of $\mathcal{V}$ \eqref{dteikon} at the beginning . Thus, \eqref{aneeikon} is indeed the final result of ANE from this method. At first, this result does not seems to perfectly match the result from point-splitting method \eqref{ane}. However, figure \eqref{fig:aneeikon} shows that the above result actually agrees with the previous method. The negative energy shock wave renders the signal to traversing the wormhole.

\begin{figure}[htbp]
    \centering
    \includegraphics[scale=0.6]{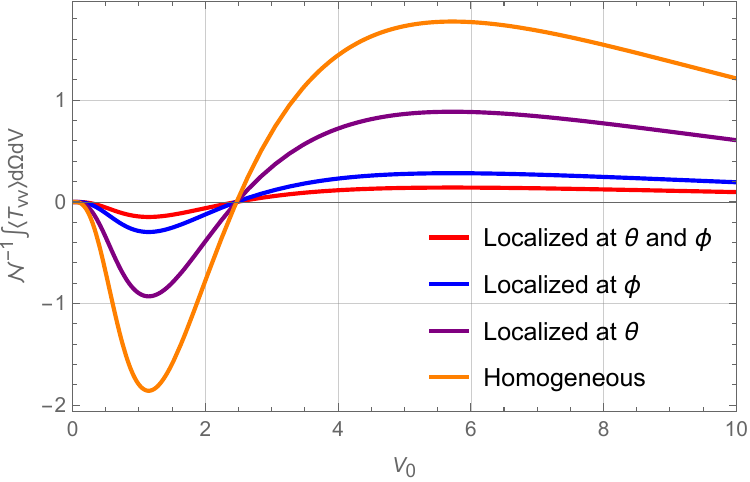}
    \caption{Average null energy from eikonal method with different choice of signal.}
    \label{fig:aneeikonsignal}
\end{figure}

We also notice that the signal choices determines the ANE's scale. Above result agrees with point-splitting method when the signal is only localized at $\phi$. For homogeneous shock wave \eqref{swtvv}, we explore other forms of signal in figure \eqref{fig:aneeikonsignal}. Homogeneous signal has the highest traversability, while fully localized signal produced the lowest one.

\subsection{Bound on Information Transfer}
After opening the wormhole by fermionic double-trace deformation, we can send information from one side to the other side. Each time we send information such as particles, the backreaction of each particle $\Delta U_{\text{each}}$ will reduce the wormhole opening $|\Delta U|$. Eventually, after some quantity of transferred information, the wormhole will be closed. We can calculate the upper bound of how much information can be transferred to the wormhole \cite{caceres2018, Freivogel2019, Ahn}. The number of information $N_{\text{bits}}$ that can be transferred is upper bounded by their total momentum $p_{\text{tot}}$ and the wormhole opening
\begin{equation}\label{upbound1}
    N_{\text{bits}}\lesssim q^{\text{tot}}_U|\Delta U|.
\end{equation}
Consider that we send a particle along the horizon $U=0$ in the form of 
\begin{equation}
    T_{UU} = \frac{q^{\text{tot}}_U}{2JL}\delta(U),    
\end{equation}
where $L=4\pi$ is the integral of the transverse coordinates, so that the total momentum of the particles is
\begin{equation}
    q^{\text{tot}}_U=2J\int d\mathbf{x} dU T_{UU}.
\end{equation}
The shock wave can also be written in terms of its total momentum $p^{\text{tot}}_V=Lp_V$. To 'fit' the wormhole, the backreaction of each particle needs to be small $\Delta U_{\text{each}}\ll1$, this probe approximation corresponds to
\begin{equation}
    \delta=\frac{4\pi G_\text{N}r_+}{4J^2L\kappa_+}p^{\text{tot}}_Vq^{\text{tot}}_U\ll1,
\end{equation}
and it starts to break down approximately at 
\begin{equation}\label{probelimit}
    q^{\text{tot}}_U\lesssim\frac{4J^2L\kappa_+}{4\pi G_\text{N}r_+p^{\text{tot}}_V}.
\end{equation}
Combining (\ref{probelimit}) and (\ref{deltaU}) into (\ref{upbound1}), the upper bound becomes
\begin{equation}
    N_{\text{bits}}\lesssim \frac{2JL}{p^{\text{tot}}_V}\left|\int d\Omega dV T_{VV}\right|.
\end{equation}

The upper bound is proportional to $J$, which near the extreme limit is $J\approx r_+^2$, consistent with the result in \cite{Freivogel2019} where the upper bound is proportional to $r_H^{d-1}$. If we take $p^{\text{tot}}_V\sim\mathcal{O}(1)$, the bound is given by the ANE and the volume of transverse space \cite{caceres2018}. The dependence on angular momentum suggests that, by adding rotation to the wormhole, we can increase the upper bound. This is equivalent as increasing black hole entropy, since we increase the black hole radius $r_+$. However, adding rotation is not enough to increase the upper bound from $\sim J$ to the theoretical maximum, which scales with the black hole entropy $S_{BH}\sim J/G_N$. We can instead increase the number of non-locally coupled fields $K$ since (without normalization) the ANE $\sim hK$ \cite{Maldacena2017,Freivogel2019,Ahn:2020csv}. The value of $K$ is not arbitrary large but upper bounded by the finiteness of black hole entropy, $K\lesssim1/G_N$. However, extremality still reduce the ANE and consequently reduce the upper bound. Furthermore, the wormhole is completely close at the extreme limit, $T_R=0$, so there is no information that can be transferred.

\subsection{Echoes Time of Wormholes Cannot Exceed Scrambling Time}
So far, we have constructed a traversable wormhole from Kerr black holes using fermionic double-trace deformation. Since the wormholes connect two different regions of Kerr black holes, we now have a symmetric effective potential system where one potential is the usual angular momentum barrier of Kerr black hole, and another potential bump is from the other side connected by the wormhole \cite{Bueno2018}. Some modes of perturbation from one boundary can pass through this potential barrier, enter the wormhole, and then be reflected by the potential from the other side. These modes are trapped between the two potentials and eventually leak out by quantum tunneling as echoes. Let us assume that the distance between the potentials is $d$, the time needed from one potential to the other and back is
\begin{equation}
    \Delta\tau = 2d.
\end{equation}
The perturbation will turn back at a certain turning point of the effective potential. The exact value of this point will not be important because we can always set them finite, especially in low frequency regime \cite{Saraswat_2020}. Therefore, we can approximate $d$ with the leading term that is the throat of the wormhole, $\Delta\tau=2|\Delta U|$. In nNHEK coordinates, we have from (\ref{kruskalkerrnex}) for certain ($V,\theta,\phi$) that
\begin{equation}
    t= \pm\frac{1}{4\pi T_R}\ln(\pm U),
\end{equation}
where $(+)$ sign means the left region and $(-)$ sign means the right region. For a perturbation that travels a distance of $|\Delta U|$ in each region, or equally left the left region at $|\Delta U|$ and then arrived at $-|\Delta U|$ in the right region, we have the time delay between echoes as
\begin{equation}\label{techo}
    \Delta t_{\text{echo}}=-\frac{1}{2\pi T_R}\ln\left(|\Delta U|\right).
\end{equation}

\begin{figure}[htbp]
    \centering
    \includegraphics[scale=0.7]{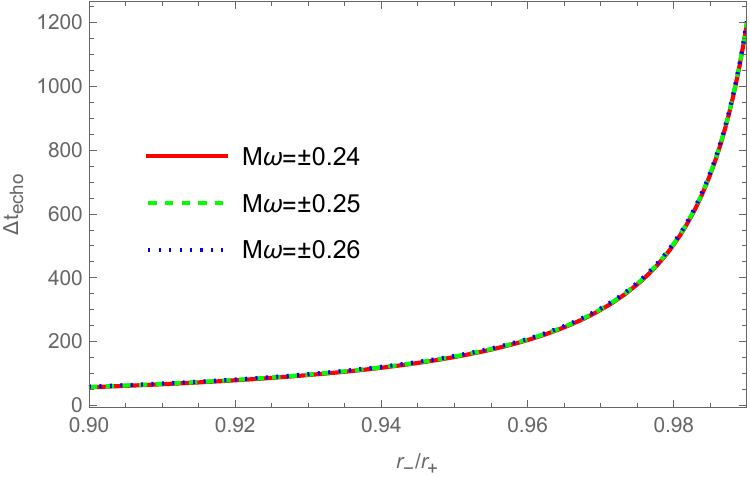}
    \caption{Time delay between echoes near the extreme limit with $V_0=1$.}
    \label{fig:echotime}
\end{figure}

Figure (\ref{fig:echotime}) shows the time delay between echoes near the extreme limit. In terms of frequency, the time delay does not show any significant differences since we vary them around the superradiant bound, so their wormhole opening is not that different either. At extreme limit, the echo time delay is divergent as the wormhole is closed, $\Delta U=0$. There are no echoes produced in that case, since the system acts like the usual black hole. 
\\
\indent The upper bound to the amount of $N_{\text{bits}}$ shown in \eqref{upbound1} can therefore give us a direct consequence to the echo time delay $\Delta t_{\text{echo}}$. By using the bound in \eqref{upbound1} with the probe limit \eqref{probelimit} and relating them with \eqref{techo}, we obtain that the echo time delay is bounded as
\begin{equation}
    \Delta t_{\text{echo}}\lesssim\frac{1}{2\pi T_R}\ln\bigg(\frac{q^{\text{tot}}_U}{N_{\text{bits}}}\bigg)\lesssim\frac{1}{2\pi T_R}\ln\bigg(\frac{J}{G_N}\frac{1}{p^{\text{tot}}_V}\left(\frac{r_+-r_-}{r_+}\right)\bigg).
\end{equation}
 The last inequality is obtained when we consider a wormhole that can at least send one bit of information, $N_{\text{bits}}\sim 1$. Note that for $p^{\text{tot}}_V\sim \mathcal{O}(1)$, since $\frac{J}{G_\text{N}}$ is in the order of the black hole entropy $S_{\text{BH}}$, we obtain
 \begin{equation}
     \Delta t_{\text{echo}}\lesssim\frac{1}{\lambda_R}\ln(S_{\text{BH}})=t_*,
 \end{equation}
 where $\lambda_R\equiv2\pi T_R$ is the associated quantum Lyapunov exponent of the black hole with temperature $T_R$ \cite{Maldacena_2016}. Therefore, we show that the echo time delay cannot exceed the scrambling time of the black hole $t_*$, which is logarithmically proportional to $S_{\text{BH}}$ \cite{Shenker_2014,Leichenauer2014}. The bound is saturated when the wormhole can only send around $\sim\mathcal{O}(1)$ bits of information.

\section{Discussions}\label{sec:discuss}
We constructed the fermionic double-trace deformation in the Kerr background using the Kerr/CFT correspondence (\ref{dtdeform}). The deformation is calculated from the boundary action in Lorentzian signature because we do not have the Euclidean near-NHEK metric. Since the only contribution at the boundary comes from the boundary term, one can add a different boundary term as long as the variational principle is upheld. In this work, we added the boundary term from \cite{Becker:2012vda} that looks like the boundary term of non-relativistic CFT. This connection may hint us at the relation between Kerr/CFT and non-relativistic conformal field theory, and needs to be studied more.

Fermionic double-trace deformation is used to modify the wormhole stress energy tensor. By considering the deformation as a perturbation, we constructed a modified two-point function that composes the stress energy tensor with point-splitting method. To calculate the two-point function, we implicitly choose a quantum vacuum state. The key physical difference of the fermionic case is that we can have a well-defined vacuum state throughout the nNHEK spacetime \cite{Casals2013}, whereas the bosonic case experiences superradiant amplification that makes the global vacuum irregular except on the axis of rotation \cite{ottewill2000, KAY199149, Bilotta}. The lack of fermionic superradiance allows us to construct TFD as the boundary state and have a consistent, well-defined wormhole geometry even at the off-axis region. This resolves one of the main difficulties encountered in the scalar case \cite{Bilotta}, where superradiant instabilities obstruct a global description. 

When calculating the modified two-point function, we choose a delta function coupling parameter (\ref{couplingh}) so that the deformation is activated at $t=t_0$. Inside the two-point function, we calculated the bulk-to-boundary propagator using the solution of Dirac equation in Kerr background. The solution is separable between the radial and angular parts. Hence, we can factor out the $(\theta,\phi)$ dependence during the calculation and then add it later at the end. A notable feature of the resulting bulk-to-boundary propagator is that the associated stress-energy tensor retains an explicit dependence on the insertion time $t_0$ even at late times. Unlike vector \cite{Ahn} and tensor \cite{khairunnisa2025} cases, we do not need to apply some prescription or ansatz previously used in \cite{Cheng_2021} to include it. This difference may be attributed to the fact that our analysis does not rely on a hydrodynamic (low energy) limit.

\begin{figure}[htbp]
    \centering
    \includegraphics[scale=0.7]{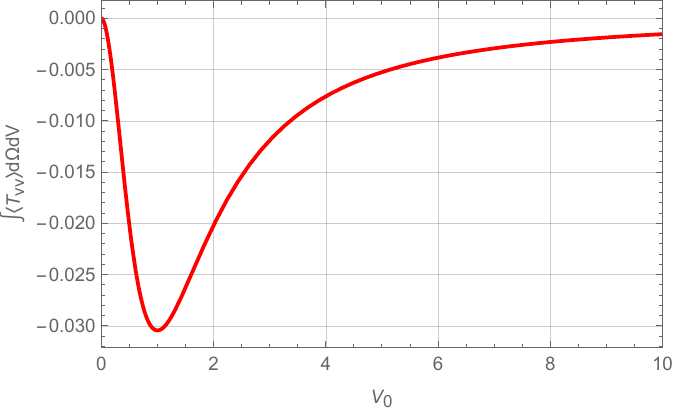}
    \caption{Average null energy when we force $m=0$.}
    \label{fig:ANEm=0}
\end{figure}

From the modified two-point function, we calculated the average null energy of the wormhole. Figure (\ref{fig:ane}) shows that the modes with angular momentum number $j=\frac{1}{2}$ dominate the ANE at the end, since the infinite summation of $j$ is convergent. The resulting time dependence of the ANE is similar to the  bosonic case: it is most negative at early insertion times, allowing traversability, while for late insertion times the wormhole becomes less traversable and eventually non-traversable. However, the key difference that we find is that the wormhole is fluctuating between traversable and non-traversable at late times. The fluctuation comes from $\Im{Q(V,V_0)}$ and $\Re{Q(V,V_0)}$ factors, while the damping comes from $P(V,V_0)^{-2-\beta}$ factors. If we write $Q(V,V_0)$ and $P(V,V_0)^{-2-\beta}$ as
\begin{equation}
    Q(V,V_0)=\frac{1}{\sin(\pi(\beta+im))}e^{im\ln\left(\frac{V/V_0-1}{1+VV_0}\right)}, \quad P(V,V_0)^{-2-\beta}=e^{-(2+\beta)\ln(P(V,V_0))},
\end{equation}
we can say that the azimuth number $m$ determines the oscillation, while $\beta$ determines the damping. In the scalar case in the Kerr background \cite{Bilotta}, the choice $m=0$ is made to avoid bosonic superradiance, which consequently restricts the wormhole to the on-axis region. In this work, we have $j=\frac{1}{2}$ as the dominant term, so we have $m=\pm\frac{1}{2}$ modes that contribute to wormhole opening. The fact that nNHEK only allows momentum around superradiant bound $\omega\simeq m\Omega_H$ and fermion does not experience superradiance, suggests that even for the lowest momentum allowed we always have fluctuating energy coming from the perturbation itself. In fact from figure (\ref{fig:ANEm=0}), if we force to take $m=0$, we get an ANE that peaked at $V_0=1$ or $t_0=0$ then less traversable at late times, similar result as in bosonic case.
Despite the fluctuating behavior, the wormhole is still non-traversable at $t_0\to\infty$ and we can set the damping by setting $\beta$ or equivalently conformal weight $h_R$.

We can set the size of the wormhole opening by fixing its frequency $M\omega$, temperature $4\pi T_R$, and fermion mass $\mu$. From figure (\ref{fig:anecaw}), although not significant because we are limited by the superradiant bound, a higher frequency results in a more negative ANE or larger wormhole opening. Together with the angular momentum number $j$ and the azimuth number $m$ for each $j$, the frequency is related to the factor $\beta$ through the separation constant. This factor determines the scaling behavior of the boundary spinor, which as mentioned previously correlates with the conformal weight in the CFT perspective \cite{Hartman:2009nz}. Fixing the conformal weight to set the wormhole opening has been done by previous work such as \cite{GaoJW,Bilotta}. In terms of temperature, a lower temperature results in a less negative ANE or a smaller wormhole opening. At the extreme limit, the wormhole is closed because there is no excitation $T_H=0$ similar to the case of a rotating BTZ \cite{caceres2018}. This is also consistent with the fact that entanglement correlation function is destroyed in the extreme limit \cite{Leichenauer2014, Tello_2023}. Therefore, there is no wormhole connecting two asymptotic regions in the ER=EPR perspective. From figure (\ref{fig:anecmu}), the ANE is less negative for a massive fermion since the mass contributes to the positive energy. While ($\omega,j,m$) are limited to fix a certain conformal weight we want, mass is the variable that we can adjust more freely, as long as we can obtain the corresponding mass-dependent separation constant.

We also investigate the traversability with eikonal method. This technique involves computation of out-of-time order correlation functions similar to those in quantum chaos \cite{Shenker:2014cwa}. It helps us to have a deeper understanding of the quantum information interpretation \cite{Maldacena2017,Gao2021Traversable,Jafferis2022Traversable}. In this perspective, the signal is scattered by the negative energy shock wave from the boundary deformation. To compare the result with point-splitting method, we take homogeneous shock wave in the transverse coordinate and localized signal at $\phi=\tilde{\phi}_1$. The scattering amplitude is approximated using the eikonal approximation. In the probe limit, when the backreaction from the signal is small, the shock wave shifts the signal trajectory in $U$ direction and the result agrees with point-splitting method.  

The wormhole opening eventually will be closed after we send information through the wormhole due to backreaction of each information. We calculated the upper bound, which is proportional to the angular momentum $J\approx r_+^2$ near the extreme limit. By increasing the rotation and the number of fields involved, we can increase the upper bound up to the scale of black hole entropy, consistent with the result from \cite{Maldacena2017, Freivogel2019, Ahn:2020csv}. However, near-extremality the wormhole is still suppressed, and at the extreme limit no information can be transferred since the wormhole is closed.

Constructing a wormhole from the Kerr black hole gives us the possibility to calculate some potential observable. One of them is the echoes \cite{Cardoso2016, Wang2018, Conklin2018, Oshita2019} that form due to the existence of symmetric potential bumps connected by the wormhole \cite{Bueno2018,pattersons2026}. The time delay between echoes is directly related to the wormhole opening logarithmically (\ref{techo}). At the extreme limit, the wormhole is closed so there are no echoes produced. The wormhole returns to the usual Kerr black hole and all of the signal is absorbed, where a similar result can be seen at a Kerr-like exotic compact object near the extreme limit \cite{Djogama2024}. The analysis of echoes can be extended by reconstructing the waveform of the echoes \cite{Bueno2018,Testa2018, Maggio2019}.

We show that the echo time delay produced by GJW traversable wormhole cannot exceed the scrambling time \cite{Shenker_2014,Leichenauer2014} of the black hole, which scales logarithmically as the black hole entropy, i.e. $\Delta t_{\text{echo}}\lesssim t_*\sim\frac{1}{2\pi T_R}\ln(S_{\text{BH}})$\footnote{The scrambling time for rotating Kerr-AdS black holes are also studied holographically in \cite{Malvimat_2022,Malvimat_2023,Malvimat_2023b,Prihadi_2023,Prihadi_2024}}. This bound tells us that \textit{a GJW traversable wormhole that can at least send one bit of information cannot produce echoes with time delay exceeding the black hole scrambling time}. Saturation occurs when the wormhole is small enough to send a single bit of information, and the signal starts to backreact on the background geometry. This can also have an observational consequence. A future detection of an echo time delay longer than $t_*$ would possibly imply that the echoes emerges from a classical noise and can be excluded from a GJW-like traversable wormhole. A relation between echo time delay and scrambling time has been studied earlier in \cite{Saraswat_2020}. In this work, we show that a similar relation also arises in the context of echoes in GJW traversable wormholes. Furhtermore, as in \cite{Ahn, khairunnisa2025}, we show that for a delta function source at late insertion time $t_0$, the wormhole opening behaves as $\Delta U\sim e^{-2\pi T t_0}$ for non-diffusive propagating probes. Therefore, the bound $\Delta t_{\text{echo}}\leq t_*$ along with \eqref{techo} also implies that the insertion time of the double-trace deformation is also bounded as $t_0\leq t_*$. This means that, a double-trace deformation that is turned on after scrambling time cannot render the wormhole traversable, as the wormhole opening size is exponentially suppressed and insufficient to transmit even a single bit of information.

The construction of a traversable wormhole with double-trace deformation has been successfully achieved using scalar \cite{GaoJW}, vector \cite{Ahn}, tensor \cite{khairunnisa2025}, and now spinor fields. The advantage of using double-trace deformation is its holographic nature, which supports the ER=EPR relation. In the future, we can try to construct a quantum teleportation protocol based on fermionic double-trace deformation, similar to the protocol that has been constructed with scalar fields \cite{Gao2021Traversable}, that might be more compatible with fermionic SYK model. We might find the difference between the bosonic protocol and the fermionic protocol due to their fundamental differences such as the symmetric vs anti-symmetric in their commutator. Another interesting aspect to study is the various black hole background where the wormhole is constructed. One of them is the Kerr-Newmann black hole that might let us combine fermion and vector double-trace deformation. It is interesting to see how these two fields interact and affect the traversability of the wormhole. We leave this for future works.

\acknowledgments
MZD would like to thank Institut Teknologi Bandung for financial support through the APS3 program. FK would like to thank the Ministry of Primary and Secondary Education (Kemendikdasmen) for financial support through the Beasiswa Unggulan program. HLP would like to thank Badan Riset dan Inovasi Nasional (BRIN) for financial support through the Postdoctoral program. Research of FPZ is supported by PPMI FMIPA ITB and the Ministry of
Higher Education, Science, and Technology (Kemendiktisaintek).

\appendix
\section{Gamma Matrices}\label{NPform}
The NP null tetrad of nNHEK is
\begin{align}
    &l^\mu = \frac{1}{r(r+4\pi T_R)}(1,r(r+4\pi T_R),0,-(r+2\pi T_R)), \nonumber\\
    &n^\mu = \frac{1}{4J\Gamma(\theta)}(1,-r(r+4\pi T_R),0,-(r+2\pi T_R)),\\
    &m^\mu = \frac{1}{2\sqrt{J\Gamma(\theta)}}(0,0,1,i\Lambda^{-1}(\theta)),\nonumber
\end{align}
satisfy the normalization and orthogonal condition $l\cdot n = -m\cdot m^*=-1$. Here we use flat gamma matrices in Weyl representation
\begin{equation}
    \gamma^0=\left(\begin{matrix}
        0 & I_{2\times2}\\
        I_{2\times2} & 0
    \end{matrix}\right),~~~~
    \gamma^i=\left(\begin{matrix}
        0 & -\sigma_i\\
        \sigma_i & 0
    \end{matrix}\right),
\end{equation}
where $\sigma_i$ are Pauli matrices and obey $\{\gamma^a,\gamma^b\}=2\delta^{ab}$. The gamma matrices in the bulk space-time are
\begin{equation}
    \gamma^\mu=\sqrt{2}\left(\begin{matrix}
        0 & \sigma^\mu\\
        \tilde{\sigma}^\mu & 0
    \end{matrix}\right),~~~~
    \sigma^\mu=\left(\begin{matrix}
        l^\mu & m^\mu\\
        m^{*\mu} & n^\mu
    \end{matrix}\right),~~~~
    \tilde{\sigma}^\mu=\left(\begin{matrix}
        n^\mu & -m^\mu\\
        -m^{*\mu} & l^\mu
    \end{matrix}\right).
\end{equation}
Using this relation, we have $\gamma^r$ for nNHEK as
\begin{align}
    \gamma^r &= \sqrt{2}\left(\begin{matrix}
        0 & 0 & 1 & 0\\
        0 & 0 & 0 & -\frac{r(r+4\pi T_R)}{4J\Gamma}\\
        -\frac{r(r+4\pi T_R)}{4J\Gamma} & 0 & 0 & 0\\
        0 & 1 & 0 &0
    \end{matrix}\right)\nonumber\\
    &= -\sqrt{2}\frac{r(r+4\pi T_R)}{8J\Gamma}(\gamma^0+\gamma^3)+\frac{\sqrt{2}}{2}(\gamma^0-\gamma^3).
\end{align}

We also choose the NP null tetrad in Kruskal coordinate as the following
\begin{align}
    &l^\mu = (0,1+UV,0,U), \nonumber\\
    &n^\mu = \frac{1}{4J\Gamma(\theta)}(1+UV,0,0,-V),\\
    &m^\mu = \frac{1}{2\sqrt{J\Gamma(\theta)}}(0,0,1,i\Lambda^{-1}(\theta)),\nonumber
\end{align}
and the corresponding covariant basis
\begin{align}
    &l_\mu = -\frac{4J\Gamma(\theta)}{1+UV}(1,0,0,0), \nonumber\\
    &n_\mu = -\frac{1}{1+UV}(0,1,0,0),\\
    &m_\mu = \sqrt{J\Gamma(\theta)}(\frac{i\Lambda(\theta)V}{1+UV},\frac{i\Lambda(\theta)U}{1+UV},1,i\Lambda(\theta)),\nonumber
\end{align}
Based on this basis, the gamma matrix $\gamma_V$ is
\begin{align}
    \gamma_V &= \frac{\sqrt{2J\Gamma(\theta)}\Lambda(\theta)U}{1+UV}\left(\begin{matrix}
        0 & 0 & 0 & i\\
        0 & 0 & -i & -\frac{1}{\sqrt{J\Gamma(\theta)}\Lambda(\theta)U}\\
        -\frac{1}{\sqrt{J\Gamma(\theta)}\Lambda(\theta)U} & -i & 0 & 0\\
        i & 0 & 0 &0
    \end{matrix}\right) \nonumber\\
    &= -\frac{\sqrt{2}}{2}\frac{1}{(1+UV)}(\gamma^0+\gamma^3)+\frac{\sqrt{2J\Gamma(\theta)}\Lambda(\theta)U}{1+UV}\gamma^2.
\end{align}
The spinor affine connection can be calculated using tetrad formalism
\begin{equation}
    \Gamma_\mu=\frac{1}{4}\omega_{(ab)\mu}\gamma^a\gamma^b, \quad \omega_{(ab)\mu}=e^\nu_b(\partial_\mu e_{a\nu} - \Gamma^\sigma_{\mu\nu} e_{a\sigma}),
\end{equation}
where tetrad from the NP null tetrad is $e^\mu_a=\frac{1}{2}((l^\mu+n^\mu),-(m^\mu+m^{*\mu}),-i(m^\mu-m^{*\mu}),-(l^\mu-n^\mu))$. Thus, $\Gamma_V$ is
\begin{align}
     \Gamma_V = \frac{1}{4}\omega_{(ab)V}\gamma^a\gamma^b = &\frac{U(4-3\Lambda^2(\theta))}{16(1+UV)}\gamma^0\gamma^3 + \frac{\Lambda(\theta)}{8\sqrt{J\Gamma(\theta)}}(1-\frac{UV}{1+UV})\gamma^2(\gamma^0+\gamma^3) \nonumber\\
     & + \frac{\sqrt{J\Gamma(\theta)}\Lambda(\theta)}{2}\frac{U^2}{1+UV} \gamma^2(\gamma^0-\gamma^3).
\end{align}

\bibliographystyle{JHEP}
\bibliography{biblio.bib}
\end{document}